\documentclass[a4paper,11pt]{article}

\usepackage[a4paper,left=2.73cm,right=2.7cm,top=3cm,bottom=3.5cm]{geometry}
\usepackage{amsmath,amssymb,hyperref,graphicx}
\numberwithin{equation}{section}

\def\be{\begin{equation}}
\def\ee{\end{equation}}
\newcommand{\bea}{\begin{eqnarray}}
\newcommand{\eea}{\end{eqnarray}}

\def\AR{A_{1}}

\def\cA{{\cal A}}
\def\cJ{{\cal J}}

\begin{document}

\par\hfill{IFUP-TH/2013-09}
\par\hfill{ICCUB-13-066}
\vspace{1cm}

\begin{center}
{\huge{\bf Charged D3-D7 plasmas: novel solutions, extremality and stability issues}}
\end{center}

\vskip 10pt

\begin{center}
{\large Francesco Bigazzi$^{a}$, Aldo L. Cotrone$^{b}$ and Javier Tarr\'\i o$^{c}$}
\end{center}

\vskip 10pt
\begin{center}
\textit{$^a$ INFN, Sezione di Pisa; Largo B. Pontecorvo 3, I-56127 Pisa, Italy.}\\
\textit{$^b$ Centro Studi e Ricerche E. Fermi, Piazza del Viminale 1, I-00184 Roma, Italy and Dipartimento di Fisica Teorica, Universit\`a di Torino and I.N.F.N. - sezione di Torino, Via P. Giuria 1, I-10125 Torino, Italy.}\\
\textit{$^c$ Departament de F\'\i sica Fonamental and Institut de Ci\`encies del Cosmos, Universitat de Barcelona, Mart\'\i\  i Franqu\`es 1, ES-08028, Barcelona, Spain.}\\

\vskip 10pt
{\small fbigazzi@pi.infn.it, cotrone@to.infn.it, j.tarrio@ub.edu}
\end{center}

\vspace{25pt}

\begin{center}
 \textbf{Abstract}
\end{center}

\noindent We study finite temperature ${\cal N}=4$ Super Yang-Mills (and more general gauge theories realized on intersecting D3-D7 branes) in the presence of dynamical massless fundamental matter fields at finite baryon charge density.
We construct the holographic dual charged black hole solutions at first order in the flavor backreaction but exact in the charge density.
The thermodynamical properties of the dual gauge theories coincide with the ones found in the usual charged D7-probe limit and the system turns out to be thermodynamically stable. 
By analyzing the higher order correction in the flavor backreaction, we provide a novel argument for the un-reliability of the charged probe approximation (and the present solution) in the extremality limit, i.e. at zero temperature.

We then consider scalar mesonic-like bound states, whose spectrum is dual to that of linearized fluctuations of D7-brane worldvolume fields around our gravity backgrounds. In particular we focus on a scalar field saturating the Breitenlohner-Freedman bound in the flavorless limit, and coupled to fields dual to irrelevant operators. By looking at quasinormal modes of this scalar, we find no signals of instabilities in the regime of validity of the solutions.

\vfill 

\newpage

\tableofcontents

\section{Introduction}

The  study of the Quantum Chromodynamics (QCD) phase diagram requires a full comprehension of finite temperature ($T$) and baryon chemical potential ($\mu$) regimes which are often beyond the reach of current theoretical methods. 
Standard perturbative treatments of QCD only provide access to certain asymptotic corners of the $(T, \mu)$ plane. Lattice simulations are limited by the so-called sign problem: a finite value of $\mu$ leads to unreliable Monte Carlo simulations (though recent advances, e.g. with analytic continuations to imaginary potential, are partially improving the situation). Moreover, since they are based on a reformulation of QCD on discretized Euclidean spaces, lattice methods are not well suited to study real-time processes.

A challenging phase of QCD is the Quark-Gluon Plasma (QGP), in which the nuclear matter dissociates into its fundamental components: quarks and gluons. 
This state of matter is created in laboratories like RHIC and LHC by collisions of heavy nuclei. 
The dynamics of the QGP is described by strong-coupling interactions and involve a finite (though typically small if compared with the plasma temperature) quark density. Moreover, most of its interesting properties reside in its real-time dynamics (hydrodynamic transport coefficients, quenching of probes moving through it, etc). Standard theoretical methods are therefore not exhaustive to comprehend its features.

Another particularly interesting part of the QCD phase diagram involves a ``large'' chemical potential (at least comparable to the zero charge crossover temperature) at zero and finite temperature.
This is the region where a critical point, if it exists, should reveal itself.
Due to its strongly coupled and large $\mu$ nature, this region of the phase diagram is beyond any current first-principle theoretical exploration.

A tool that has been used long and wide in the last years to understand the dynamics of strongly coupled quantum field theories is the gauge/gravity (holographic) correspondence. This allows to map classes of strongly coupled quantum field theories into weakly coupled theories of gravity in at least one dimension more. The extra (radial) direction is the geometric counterpart of the Renormalization Group (RG) scale of the dual field theory. 

The best understood realization of the correspondence, known as AdS/CFT, relates gravity theories on asymptotically Anti-de-Sitter ($aAdS$) backgrounds in $d+1$ spacetime dimensions to conformal field theories in $d$-dimensional Minkowski spacetime.  
Finite temperature and chemical potential phases in these theories are then mapped into dual $aAdS$ charged black hole backgrounds.

Although there are several examples in which the correspondence between a quantum field theory and a dual theory of gravity is explicitly realized, this is unfortunately not the case for QCD: the holographic dual of Quantum Chromodynamics remains out of reach.
Maximally supersymmetric (${\cal N}=4$) $SU(N_c)$ Yang-Mills (SYM) theory in four dimensions, instead, has a well known holographic dual, and most of its non-perturbative properties can be exactly deduced using the correspondence. It is a conformal field theory with a bosonic content of six scalars and one vector in the adjoint representation; the scalar manifold has $SO(6)$ symmetry. 
Despite the enormous differences among this theory and QCD, in the last years we have learned that it provides sensible benchmarks on some equilibrium and dynamical features of phases where QCD could be roughly approximated by a (strongly coupled) conformal model. This is the case of the deconfined QGP phase, where e.g. the trace anomaly results to be quite suppressed for not too large values of the temperature above the confinement-deconfinement crossover. On the other hand, turning on temperature badly breaks supersymmetry, thus allowing a phenomenologically unpleasant feature of the ${\cal N}=4$ SYM model to be cast away. See \cite{matrev} for a recent review on all these issues.

A QCD-inspired refinement of the above toy model requires at least the inclusion of  quarks, i.e. fermionic matter fields transforming in the fundamental representation of the gauge group. Actually, in order to preserve some supersymmetry in the $T=0$ vacuum (so to benefit from a quantitative control on the related gravity dual) one has to introduce flavor multiplets, containing scalar (s-quarks) fields as well. The inclusion of these multiplets generically breaks part of the global symmetries of the original model. The case we will focus on in this paper is that of massless fundamental hypermultiplets which, in the ${\cal N}=4$ SYM case, are introduced in such a way to preserve an $SO(4)\times U(1)_R\subset SO(6)$ symmetry subgroup, where $U(1)_R$ is an R-symmetry. 

The addition of fundamental degrees of freedom also breaks conformal invariance, though in an unpleasant way. The beta function for the 't Hooft coupling $\lambda=g_{YM}^2 N_c$, results to be positive (thus signaling the occurrence of a UV Landau pole) and proportional to the number of flavored species, $\beta(\lambda) \sim (N_f/N_c)\lambda^2$, as opposed to the negative beta function of QCD. Just as in the case of  QED, another famous theory exhibiting a UV Landau pole, a sensible treatment of the model can be obtained only by focusing on the IR physics.

The present work will actually focus on ${\cal N}=4$ SYM theory (and its ``quiver'' generalizations thereof, see below) coupled with massless dynamical flavors. We will discuss its behavior, in a limiting regime of its parameters, at finite temperature and quark chemical potential (with both scales taken to be much below the UV Landau pole), by means of a dual charged black hole background (the case of extremely massive flavors has been analyzed in \cite{Kumar:2012ui}). The gravity solution presented in this paper represents an improvement with respect to the ones found in the past by the authors and collaborators, in that it allows to scan an enlarged region of the $(T,\mu)$ plane. 
To be specific, our solution will be at first order in the flavor backreaction parameter $\epsilon \sim \lambda N_f/N_c$, but exact in the chemical potential.
So, we will be \emph{formally} allowed to explore the large charge regime of the system.

As a first application, we will verify that the backreacted solution reproduces the thermodynamics calculated in the probe approximation \cite{Kobayashi:2006sb} (in particular, the system turns out to be thermodynamically stable).
This is a consistency check that the solution correctly describes the same physics as the probe whenever observables are computable in both frameworks.
That is, the backreacted solution faithfully contains and extends the probe physics.

Moreover, we will be able to check the consistency of the large charge regime in the above model as well as in more general ``D3-D7 systems'' (see the technical overview below).
As we will see, in the parametrically large charge regime (including the zero temperature case),\footnote{With ``parametrically large charge'' we mean that the charge or chemical potential is larger by factors of $N_c$ than the other energy scales in the problems, as e.g. the temperature (this is the case explored in this paper), the flavor masses or possibly a dynamically generated IR scale (as in confining YM theories, such as the Sakai-Sugimoto model \cite{Sakai:2004cn}). In particular, the extremely interesting case of zero temperature conformal charged systems with massless matter falls in this definition.} higher corrections in the backreaction of the flavors, that is in $\epsilon$, are \emph{not} subleading, spoiling the reliability of the solution at leading order in $\epsilon$.
As a byproduct, this means that \emph{the probe approximation is not reliable in the IR at parametrically large charge or chemical potential}.
This fact is known, see e.g. \cite{Hartnoll:2009ns}, but possibly not fully appreciated. 
For example, it means that the use of probes to study the IR regime of the zero temperature, zero flavor mass limit of charged conformal systems is not trustworthy.
Here we are able to confirm this fact from the explicit analysis of the brane backreaction.

Finally, we will start investigating whether flavor dynamics affects the stability of the system. Our analysis extends previous results in the literature, where the flavors were treated as non-dynamical probes. As it was pointed out in several places (see e.g. \cite{Ammon:2011hz}), there are various instabilities one could expect in our models. Charged fermions and scalars on the field theory side can give rise to various kinds of condensates: the corresponding symmetry broken phases should be possibly accounted for by novel dual gravity solutions. Instabilities could also show up in the spectrum of uncharged mesonic-like bound states. This is particularly interesting for us since it can be determined by studying (linear) fluctuations of dual fields around our charged black hole backgrounds. Focusing on a possibly critical scalar meson subsector, we will show that no instability appears at zero momentum in the regime where our gravity solution can be trusted.

To help the reader going into the details of our analysis, we provide below a short technical overview on the holographic approach to flavored gauge theories, pointing towards the main issues raised in this paper. 
The reader familiar with the subject can safely skip (part of) this subsection.
The way this paper is organized is then presented in a further subsection.  

\subsection{Technical overview}

Holographically, in the $N_c\gg 1$, $\lambda\gg1$ limit, flavorless ${\cal N}=4$ SYM theory is described by type IIB supergravity in $AdS_5 \times S^5$. The  $S^5$ has a $SO(6)$ isometry that relates directly to the symmetry of the scalar manifold in the field theory. Turning on temperature on the field theory side is accounted for by placing a black hole at the center of $AdS_5$.
The 10d geometry originates as the near horizon solution of a stack of $N_c$ D3-branes. 
Introducing flavor into the theory is accounted for by adding a stack of suitably embedded $N_f$ D7-branes \cite{Karch:2002sh}. The latter wrap an $S^3\subset S^5$, which translates into the $SO(4)\times U(1)_R\subset SO(6)$ symmetry breaking pattern. The flavored SYM theory arises in turn as the low energy description of the open strings which end on the D3-D7 branes. Turning on a finite baryon charge density requires turning on an electric field on the D7 worldvolume.

The D7-brane dynamics is described at leading order by the Dirac-Born-Infeld (DBI) and Wess-Zumino (WZ) actions. The finite value of the beta function, mentioned above, is accounted for by a running dilaton, which blows up at a radial position holographically dual to the field theory UV Landau pole.

The treatment of the system consisting of type IIB supergravity with DBI and WZ source terms is generically a quite complicated problem to work out.
We must, therefore, work in an appropriate approximation if we want to tackle the system.\footnote{See \cite{tutti} for examples of other solutions to this problem in the supersymmetric case.}

\paragraph{Probe approximation.}

One of such approximations was started with the seminal paper of Karch and Katz \cite{Karch:2002sh}, and  goes under the name of \emph{probe (or quenched) approximation}.
It consists in considering a small number of fundamental degrees of freedom as compared to the adjoint ones: $N_f/N_c \ll 1$; then the 't Hooft coupling beta function is very small and can be effectively set to zero. This is the 't Hooft approximation.
On the field theory side this corresponds to discarding effects of the dynamics of fundamental matter in the theory, which in the language of lattice gauge theory corresponds to quenching the quarks.
On the geometrical side of the correspondence, the effect is to treat the D7-branes as probes on top of the undeformed $AdS_5\times S^5$ background. If these branes are all coincident, ${\cal N}=2$ supersymmetry is conserved and the flavor group is $SU(N_f)$. 
Since in the undeformed background the dilaton is constant, there is no RG flow, signaling that in the field theory we have completely neglected the flavor field effects on the original vanishing beta function.
As the gravitational content of the theory is kept frozen in the presence of the D7-branes, the dynamics of the flavors is described entirely by the DBI and WZ actions.

Soon after the publication of  \cite{Karch:2002sh}, in which the supersymmetric D7 probes were introduced,  the spectrum of mesonic-like bound states in the supersymmetric theory was analyzed in \cite{Kruczenski:2003be}. In that paper, the spectrum of perturbations of several D7-brane fields, and their dual gauge theory operators, were studied and organized into the appropriate supersymmetric multiplets. Later, the theory was extended to include finite temperature physics \cite{Mateos:2006nu,Mateos:2007vn} and the presence of a finite quark density or chemical potential \cite{Kobayashi:2006sb,Karch:2007br,Mateos:2007vc}. 
Following the construction of the thermal/charged backgrounds, perturbations of the system were studied in  several regimes, see e.g. \cite{Mateos:2007yp,Myers:2008cj,Mas:2008jz,Erdmenger:2008yj}.

The contribution of the charged massless hypermultiplets to the free energy, to leading order in the probe approximation, was studied in \cite{Karch:2007br}, at finite or zero temperature. As a result the thermodynamics of the model was determined, showing that the system is thermodynamically stable. 

In \cite{Ammon:2011hz} the possible occurrence of instabilities was studied by considering the mesonic spectrum at finite charge and zero temperature for general flavor mass. In that paper it was outlined how the fluctuations of all the bosonic D7-brane fields effectively probe an $AdS_2$ near-horizon metric. This ``effective'' $AdS_2$ region could be the remnant of a corresponding extremal region in the fully-backreacted background metric. In this case one would expect it to be related to a finite entropy degenerate system at $T=0$ as in the usual AdS-Reissner-Nordstr\"om (AdS-RN) solution. Such a degeneracy would then also be considered as a signal of possible instabilities (towards non-degenerate states) to occur. 

An especially interesting mesonic operator is the dimension-two scalar corresponding to a worldvolume scalar field of mass squared $m^2 L^2=-4$, with $L$ being the $AdS_5$ radius. This mass sits precisely on top of the Breitenlohner-Freedman (BF) bound for $AdS_5$, and a small perturbation could push the mode beyond unitarity. By studying the quasinormal modes of that scalar (corresponding to poles in the retarded two-point function of the dual mesonic operators), the authors of \cite{Ammon:2011hz} found that none of them corresponds to an instability. In turn, they found a ``diffusive'' mode (analogous to the ``zero sound mode'' appearing as a pole in the $U(1)_B$ charge density two-point function) with purely imaginary dispersion relation, $\Omega = - i D q^2$, with $q$ the momentum of the mode and diffusion constant $D\sim 1/\mu$ from dimensional analysis. Near the origin of $AdS_5$ spacetime, the mode is approximated by a massless scalar in $AdS_2$. Since a zero mass scalar is above the $AdS_2$ BF bound, the absence of instabilities consistently follows. 
In \cite{Ammon:2012mu} it was shown that this mode is related to the Higgs branch of a moduli space of vacua (at zero temperature and finite chemical potential).

In summary, despite the suggestions that the model should present an instability at low temperature, none could be found. Since all the relevant bosonic fields were included in the analysis of \cite{Ammon:2011hz}, it is possible that the instability can be seen only when backreaction of the D7-branes on the geometry is included.
In view of the previous claim on the validity of the probe, however, one has to take in mind that the results described above are on solid grounds only away from the zero temperature, zero flavor mass limit.

\paragraph{Smearing approximation.}

A different limiting regime, considered in the present paper, is provided by the Veneziano approximation, in which both $N_f$ and $N_c$ are taken to be large with their ratio $N_f/N_c$ fixed. In this way, the dynamics of the quarks (in a perturbative language, the quark loops) is not ignored. In the holographic context this consists in considering the backreaction of the D7-branes in the original $AdS_5 \times S^5$ geometry.

When the flavor branes are coincident, there are limitations (at least in flat space) in the maximum number we may have (see \cite{Erdmenger:2007cm} for an explanation in this context).
To avoid this and other (more technical than physical) problems, the D7-branes are distributed homogeneously in the two dimensions perpendicular to their worldvolume. This strategy was initiated in \cite{Bigazzi:2005md,Casero:2006pt}. The transverse distribution breaks part of the supersymmetry as well as the flavor group which reduces to its maximal Abelian subgroup $U(1)^{N_f}$. This is usually referred to as the \emph{smearing approximation}. 

As already stated above, ${\cal N}=4$ SYM theory in the presence of fundamental matter develops a Landau pole at a certain energy scale (governed by $N_f/N_c$). Holographically this appears because the D7-brane backreaction sources the equation of motion for the dilaton, which does not admit a constant solution any longer. Actually, the dilaton shows a logarithmic divergence  at a finite radial position $r = r_{LP}$ corresponding to the field theory UV Landau pole. In particular this implies that a conformal boundary is no longer available and the existence of a holographic dictionary is in dispute. However, experience with this and other systems suggests that physics of the IR can still be defined by applying the standard holographic dictionary at a large cutoff  radial position below $r_{LP}$.
This is the approach we follow in this paper, where we formally push the position $r_{LP}$ to infinity. 

\paragraph{D3-D7 systems in the smearing approximation.}

A ${\cal N}=1$ supersymmetric solution describing the backreacted intersection of D3 and smeared D7 branes at zero temperature and charge density was derived in \cite{Benini:2006hh}. The construction is quite general in that it is easily extended to infinite classes of flavored ${\cal N}=1$ quiver theories which arise when the D3-branes are placed at the tip of toric Calabi-Yau cones.\footnote{In the flavorless case, when these theories are conformal, the dual gravity backgrounds have $AdS_5\times X_{SE}$ metric, where $X_{SE}$ is the 5d compact Sasaki-Einstein base of the given cone.} 

The solutions in \cite{Benini:2006hh} are the starting point of the construction in \cite{Bigazzi:2009bk} where their black hole (finite temperature) versions were found. Since the models are no longer supersymmetric, the equations of motion one needs to solve are second order and the analysis was restricted to small backreaction parameter $\epsilon_*\sim \lambda_* N_f/N_c$ (where $\lambda_*$ is the 't Hooft coupling at a given scale, see the text for details) to keep analyticity. The case of the charged black hole was later studied in \cite{Bigazzi:2011it,Bigazzi:2011db} by including the effects of finite, small charge density per flavor perturbatively in the black hole solution of \cite{Bigazzi:2009bk} (which is itself perturbative in $\epsilon_*$).\footnote{These solutions have been studied further in \cite{studies1,studies2}.} In the present paper we generalize this work by giving the exact, in charge density per flavor, charged black hole solution (again, still perturbative in $\epsilon_*$).
As discussed before, we are able to use this solution to check the validity regime of the small $\epsilon_*$ approximation, and of the probe approximation as well. 

A reason to construct this solution is to study the effects of backreaction in the stability analysis of \cite{Ammon:2011hz} recalled above. To perform this analysis we need to make a consistent perturbation of the IIB+DBI+WZ equations of motion. To do this we use extensively the results of \cite{Cotrone:2012um}, where a consistent reduction of the IIB+DBI+WZ action to five dimensions was presented. In this reduction the fields appearing are: 
\begin{itemize}
\item[-] fields preserving the complex structure of the internal 5d manifold on which one compactifies (to ensure the consistency of the reduction);
\item[-] fields neutral under the $U(1)_R$ R-symmetry, which is preserved by the smearing (this condition can be relaxed, see \cite{Cassani:2010uw,Liu:2010sa,Gauntlett:2010vu}).
\end{itemize}
We reproduce table \ref{tab.fields} from \cite{Cotrone:2012um} where the field content of the 5d reduction and its 10d origin is specified. The action and equations of motion dictating the dynamics of these fields are written in \cite{Cotrone:2012um}. We will not write them here explicitly but will refer to them extensively. The field strengths follow an equivalent nomenclature to the potentials shown in the table. For example, the 10d field strength three-form, $F_3$, gives rise to three different 5d field strengths $F_1^{(3)}$ (a one-form), $F_2^{(3)}$ (a two-form) and $F_3^{(3)}$ (a three-form). We refer again to \cite{Cotrone:2012um} for exact definitions and details.
\begin{table}[tb]
  \centering
  \begin{tabular}{|c|c|c|c|c|}
    \hline
    Original 10d field & 5d two-forms & 5d vectors & 5d scalars & 5d metric \\ 
    \hline
    $C_{4}$ & $$ &  $C^{(4)}_{1}$ & $C^{(4)}_{0}$ & $$ \\
    $C_{2}$ & $C^{(2)}_{2}$ & $C^{(2)}_{1}$ & $C^{(2)}_{0}$ & $$ \\ 
    $C_{0}$ & $$ & $$ & $C^{(0)}_{0}$ & $$ \\ 
    $B_{2}$ & $B^{(2)}_{2}$ &$B^{(2)}_{1}$ & $B^{(2)}_{0}$ & $$ \\
    $\cA$ & $$ & $\cA_{1}$ & $\cA_{0}$ & $$ \\
    $G$ & $$ & $\AR$ & $f,w$ & $g$ \\
    $\Phi$ & $$ & $$ & $\Phi$ & $$ \\
    \hline
  \end{tabular}
  \caption{\label{tab.fields}Five-dimensional fields originating from the ten-dimensional ones, taken from \cite{Cotrone:2012um}. $G$ and $\cA$ represent the 10d metric and world-volume gauge field, respectively.}
  \label{tab:label}
\end{table}

The interesting scalar field sitting on the $AdS_5$ BF bound (the one  giving rise to the diffusive mode in \cite{Ammon:2011hz}) corresponds in the former classification to the scalar $\cA_{0}$ coming from the world-volume vector on the D7-branes. This scalar, in fact, is dual to an operator of dimension $\Delta=2+Q_f$,\footnote{For smeared anti-D7-branes this mode is not present, and $\cA_{0}$ describes a $\Delta=6+Q_f$ operator.}  where $Q_f$ is a measure of the backreaction (related to $\epsilon_*$ above by a factor of the dilaton, see \eqref{epsilondef} later). We could wonder whether the $\lim_{q\to0}\Omega=0$ limit, found  for the diffusive mode in the probe approximation, is perturbed in the backreacted case. Even if $\epsilon_*$ is taken to be small, it could weight positive or negative corrections to this limit, leading in one case or another  to an instability (depending on the conventions: for us purely imaginary modes with $\mathrm{Im}(\Omega)>0$ will be unstable, i.e. exponentially growing). 

\subsection{Organization and main results of the paper}
This paper is organized as follows. In section \ref{sec.chargedblackbrane} we present the minimal effective 5d gravity action relevant for the holographic study of the charged D3-D7 system with massless flavors. We will setup our perturbative approach presenting it in a slightly different (though physically equivalent) way as that followed by the authors and collaborators in the past (see ref. \cite{Cotrone:2012um}). As we will see, this approach is quite useful in that it automatically identifies the fields contributing to the effective action at each order in the $\epsilon_*$ expansion. In particular it allows one to holographically study how irrelevant operators are ``integrated'' out along the RG-flow towards the IR. 
Moreover, as we will see, it shows how the effective action, order by order, assumes quite simple forms (e.g. Einstein-dilaton, Einstein-DBI) which are commonly used in bottom-up holographic setups. Finally, the approach allows to easily recover the already known uncharged perturbative black hole solution. 
The novel content of section \ref{sec.chargedblackbrane} is a charged black hole solution at first order in $\epsilon_*$. The difference with the solution presented in ref. \cite{Bigazzi:2011it} is that the parameter related to the charge density per flavor is treated in full generality in the present work, whereas in previous solutions it was treated perturbatively. This complicates the explicit form of the solution but increases its regime of validity. 

In section \ref{sec.thermo} we study the thermodynamics of the charged solution at first order in $\epsilon_*$ and compare it to the results in \cite{Bigazzi:2011it} and with those obtained in the probe approximation. We will show, in turn, that the system is thermodynamically stable.

In section \ref{sec.extremal}, by studying second order corrections to the charged solution of section \ref{sec.chargedblackbrane}, we will show how in the perturbative-in-$\epsilon_*$ regime (and thus in particular in the probe approximation with massless flavors) it is not sensible to approach the extremal $T/\mu\rightarrow 0$ limit, for which an all-order solution is necessary. 

In section \ref{sec.perturbations} we will study linear perturbations on top of the charged solution, focusing in particular on the mode dual to the operator with dimension $\Delta=2+Q_f$. We will find no instability in the regime of parameters where the solution is reliable.

We present conclusions and final comments in section \ref{sec:conclusions}. Further useful results are collected in an appendix.

\section{The charged D3-D7 system} \label{sec.chargedblackbrane}

A solution of the reduced system described in the introduction corresponds to D7-branes with a finite electric field in the radial direction.
This describes holographically  the presence of  charge density per quark in the field theory.
The study of this setup was initiated in \cite{Bigazzi:2011it}, where it is shown that the minimal set of fields that must be considered includes three scalars, $\Phi$, $w$ and $f$, dual to operators of dimension $\Delta_\Phi=4$, $\Delta_w=6$ and $\Delta_f=8$ respectively, two vectors $\cA_1$ and $C_1^{(2)}$ and one two-form $C_2^{(2)}$, that upon redefinitions give rise to a massless vector field, a massive vector field, and a massive two-form, corresponding to a $\Delta_J=3$ flavor current operator, and $\Delta_V=2+\sqrt{9+\sigma^2 Q_f}$  and  $\Delta_{T}=6+Q_f$ operators.\footnote{A fourth scalar $C_0^{(2)}$ corresponds to a St\"uckelberg scalar coupled to the $C_1^{(2) }$ vector and can be gauged away. The redefinitions leading to these dimensionalities can be found in \cite{Cotrone:2012um}.}
The parameter $\sigma$  is equal to $-1$ (resp. to $+1$) if D7-brane (resp. anti-D7-branes) are introduced in the setup (see \cite{Cotrone:2012um}). The case we will focus on is the $\sigma=-1$ one, though some of our results are presented for the general case. 
Schematically, the operators dual to the fields present in our solution are of the form \cite{Ceresole:1999zs,Benvenuti:2005qb,Benini:2006hh,Erdmenger:2007cm}
\begin{align*}
{\cal O}_\Phi & =  \mathrm{tr} F^2  \ , \qquad
{\cal O}_f  = \mathrm{tr} F^4 \ , \qquad
{\cal O}_w  \in  \mathrm{tr} ({\cal W}_\alpha{\cal W}^{\alpha}) \ , \\
{{\cal O}_J}^\mu & =  \psi^{\dagger\alpha}\gamma^\mu_{\alpha\beta} \psi^\beta + i q^{\dagger m}D^\mu q^m-i \bar D^\mu q^{\dagger m}q^m  \ ,
\\
{\cal O}_V, {\cal O}_T & \in Tr(\bar {\cal W}_{\dot\alpha} {\cal W}_{\beta}{\cal W}^{\beta})+...    \ ,
\end{align*}
where $\cal W$ is the gluino superfield, 
$\psi$ a doublet of spinors and $q$ a doublet of squarks.

\subsection{The 5d effective action}
A 5d effective gravity action describing the dynamics of the minimal set of fields considered above can be obtained  from a more general one \cite{Cotrone:2012um} arising from a \emph{consistent} Kaluza-Klein truncation of 10d supergravity. With the specific choice of  the ansatz we will take below, the dynamics of the minimal set of fields relevant for  the charged D3-D7 solutions is described by
\begin{eqnarray}\label{5dcharged}
S_5 &=& \frac{1}{16\pi G_5} \int d^5 x \sqrt{-\det g} \left[ R[g] +{\cal L}_s  +{\cal L}_f + {\cal L}_{DBI}\right]\,,\label{eq.effective5daction} \\
{\cal L}_s&=&- \frac{40}{3}(\partial f)^2 - 20 (\partial w)^2 - \frac{1}{2}(\partial\Phi)^2 - V(\Phi,f,w)\,,\nonumber\\
V(\Phi,f,w)&=& 4 e^{\frac{16}{3}f+2w} \left( e^{10w}-6\right) + \frac{1}{2} Q_f^2 e^{\frac{16}{3} f -8w +2\Phi} + \frac{Q_c^2}{2} e^{\frac{40}{3} f}\,,\nonumber\\
{\cal L}_f &=&-\frac12e^{\Phi-\frac{4}{3}f-8w}(d {\cal C}_1 +Q_f d{\cA}_1)^2- 4 e^{\Phi+4f+4w}{\cal C}_1^2 -\frac12e^{\Phi-\frac{20}{3}f} (d{\cal C}_2)^2\,,\nonumber \\
{\cal L}_{DBI}&=&-4Q_f e^{\Phi+\frac{16}{3}f+2w}\sqrt{1+\frac12e^{-\Phi-\frac{20}{3}f}(d{\cal A}_1)^2}\ \nonumber .
\end{eqnarray}
We have redefined
\be
C_2^{(2)}\equiv {\cal C}_2\,,\quad C_1^{(2)}\equiv {\cal C}_1\,, 
\ee
and reabsorbed a $2\pi\alpha'$ factor in the definition of ${\cal A}_1$. The constant $Q_f$ is related to the number $N_f$ of D7-branes as follows
\be
%Q_c = \frac{g_s {\alpha'}^2 N_c (2\pi)^4}{V(X_{SE})}\,,\quad 
Q_f = \frac{V(X_3)}{4 V(X_{SE})} g_s N_f\,.
\ee
Here $V(X_{SE})$ and $V(X_3)$ are the volumes of the five-dimensional internal Sasaki-Einstein manifold and of the three-cycle wrapped by the D7-branes, respectively.\footnote{In the flavored ${\cal N}=4$ SYM case, for example, $V(X_{SE})=V(S^5)=\pi^3$, $V(X_3)=V(S^3)=2\pi^2$.}

The action above admits an $AdS_5$ solution when $Q_f=0$ and all the matter fields are trivial. The $AdS$ radius is related to the constant $Q_c$ by $L^4 = Q_c/4$. When choosing to work in $L=1$ units (as we will mostly do here) one has then to set $Q_c=4$ accordingly.\footnote{Reinserting the correct dimensionalities one finds $ Q_c = (g_s {\alpha'}^2 N_c (2\pi)^4)/V(X_{SE})$ where $N_c$ is the number of D3-branes.}

The 5d Newton constant $G_5$ is obtained from the 10d one after reduction on the 5d internal compact space (of radius $L$) and it is thus given by
\be
\frac{1}{16\pi G_5} = \frac{L^5V(X_{SE})}{(2\pi)^7 g_s^2 \alpha'^4} \ .
\ee

Using the action \eqref{eq.effective5daction}, care has to be given to the fact that it does not strictly arise as a consistent truncation from 10d, but just as a ``partially on-shell'' reduction of the latter on the homogeneous ``electric ansatz'' we will choose. For example there will be extra conditions to be imposed on $d{\cal C}_2$ coming from the 5d equations of motion for $H_3$ (which we take to be zero here). We will write these extra conditions explicitly, deriving them from the action given in \cite{Cotrone:2012um}, as an expression for the two-form field, i.e., we will not impose an ansatz for ${\cal C}_2$, since it will be determined from the other fields in our setup.

The black hole metric ansatz we will consider is of the form
\be
ds_5^2 = e^{2A(r)}[ - b(r) dt^2 + dx_i dx_i ] + e^{2B(r)} \frac{dr^2}{b(r)}\,.
\label{metans}
\ee
Moreover, we will assume the scalar fields $f,w, \Phi$ to be functions of the radial coordinate only.  
For the vector fields we will choose the following electric ansatz
\be
{\cal A}_1 = {\cal A}_t (r) dt\,,\quad {\cal C}_1 = {\cal C}_t (r) dt\,.
\label{formans}
\ee

\subsection{The perturbative approach}
Let us split the dilaton into a constant and a piece vanishing at the particular position $r_*$
\be
\Phi(r) = \Phi_* + \phi(r) \ , \qquad \phi(r_*) = 0 \ .
\ee
The thermal smeared D3-D7 systems studied so far (and in this paper) are perturbative in the parameter \cite{Bigazzi:2009bk}
\be\label{epsilondef}
\epsilon_* \equiv e^{\Phi_*} Q_f = \frac{V(X_3)}{16\pi V(X_{SE})} \lambda_* \frac{N_f}{N_c} \ ,
\ee
where the 't Hooft coupling $\lambda_* = 4 \pi g_s e^{\Phi_*}N_c$ is defined at a certain scale set by radial coordinate $r = r_*$. We can analogously define a running parameter $\epsilon(r)= Q_f e^{\Phi(r)}$. Since the flavor perturbation induced by the D7-branes and dual to the dilaton field is marginally irrelevant, we know (see \cite{Bigazzi:2009bk}) that the shifted dilaton is logarithmically running to leading order: $\phi(r)\sim \epsilon_* \log(r/r_*)$. This implies that, perturbatively, the beta function for $\epsilon(r)$ is proportional to $\epsilon^2$. This in turns implies that, at first order, the differences between $\epsilon_*$ and any other allowed value $\epsilon(r_0)$ are subleading in $\epsilon_*$. When working with the first order thermal solutions, whose horizon radius $r_h$ is holographically related to the field theory temperature, we will  then be free to replace $\epsilon_*$ with $\epsilon_h=\epsilon(r_h)$ at first order in the backreaction parameter.

As it was discussed in \cite{Bigazzi:2009bk}, working in the $\epsilon_*\ll1$ limit, allows us to decouple the IR physics (to which we are ultimately interested in) from the troubling UV Landau pole. Actually, the limit allows us to trust our IR description up to an arbitrary ``large'' radial cutoff $r=r_s$ (where we impose our solutions to match with the uncharged $T=0$ ones found in \cite{Benini:2006hh}), which has to be smaller than the one corresponding to the UV Landau pole scale.\footnote{Actually, it must be $r_s<r_a<r_{LP}$, where $r_a$ is the radial position where the holographic $a-$function presents a singularity \cite{Bigazzi:2009gu}.}

Our approach to the perturbative solution here will be slightly different  (though physically equivalent) to the one usually followed so far in the literature. The standard treatment consists in first writing down the exact-in-backreaction equations of motion and then expanding them in series of $\epsilon_*$. In this paper, instead, we will first expand the action and only then derive the equations of motion. This approach is particularly useful if one is interested in determining holographically the IR properties of the D3-D7 plasma, like the thermodynamics. Since the field theory partition function is obtained from the on-shell (renormalized, Euclidean) 5d action, expanding the latter in powers of $\epsilon_*$ allows us to understand which fields will actually contribute to each order.

Moreover, unless specified otherwise, we will formally send the cutoff  $r_s$ to infinity, which physically amounts in neglecting all power-like $r/r_s$  terms in our solutions. This will  allow us to automatically focus on the deep IR solutions which are those relevant for describing the thermodynamics.

Before presenting the charged perturbative solution, let us first review what happens in the uncharged case using the approach described above. 

\subsection{The uncharged case}\label{sec:unch}
Setting the $AdS$ radius to one, the effective 5d gravity action in the uncharged case reduces to  
\begin{equation}\label{5daction}
S_5 = \frac{1}{16\pi G_5} \int d^5 x \sqrt{-\det g} \left[ R[g] - \frac{40}{3}(\partial f)^2 - 20 (\partial w)^2 - \frac{1}{2}(\partial\Phi)^2 - V(\Phi,f,w) \right]\, ,
\end{equation}
where
\begin{equation}
V(\Phi,f,w) = 4 e^{\frac{16}{3}f+2w} \left( e^{10w}-6 + Q_f e^{\Phi}\right) + \frac{1}{2} Q_f^2 e^{\frac{16}{3} f -8w +2\Phi} + 8 e^{\frac{40}{3} f}\, .
\label{potV}
\end{equation}
When $Q_f=0$ the action admits an $AdS_5$ (black hole) solution of unit radius where $\Phi=$const and $f=w=0$. The perturbative solution in $\epsilon_*=Q_f e^{\Phi_*}\ll1$, will arise in form of an expansion around the unflavored $AdS_5$ background. In particular, we will expand the scalars as follows:
\bea
\Phi(r)&=&\Phi_* + \epsilon_*\, \phi_1(r) + {\cal O}(\epsilon_*^2) \ ,\nonumber \\
f(r)&=& \epsilon_* \left[-\frac{1}{40} +\sqrt{\frac{3}{80}}\,f_1(r)\right]+ {\cal O}(\epsilon_*^2)\,,\nonumber\\
w(r)&=& \epsilon_* \left[-\frac{1}{60}+\sqrt{\frac{1}{40}}\,w_1(r)\right]+ {\cal O}(\epsilon_*^2)\,.
\label{redef}
\eea
The constants in $f,w$ are chosen in order to diagonalize the action and to have canonically normalized kinetic terms. Plugging these expressions in (\ref{5daction}) and requiring that $\epsilon_* \psi_1\ll1$ for $\psi_1=(f_1,w_1,\phi_1)$, we find, to second order in $\epsilon_*$,
\be
S_5=\frac{1}{16\pi G_5}\int d^5x \sqrt{-\det g}\left[R[g] + 12 \left(1-\frac{\epsilon_*}{3}-\frac{\epsilon_*^2}{72}\right) + \epsilon_*^2\,{\cal L}_m\right]\,,
\label{seceffun}
\ee
with
\be
{\cal L}_m = - \frac12(\partial f_1)^2 - \frac12 (\partial w_1)^2 - \frac{1}{2}(\partial\phi_1)^2- v(\phi_1,f_1,w_1)\,,
\ee
and
\be
v(\phi_1,f_1,w_1) = 4\phi_1+ 16 f_1^2 + 6 w_1^2\,.
 \ee
This shows that the masses of the modes $f_1, w_1$ around $AdS$ are $m_f^2 = 32$ and $m_w^2=12$ and thus (as already pointed out in the literature) that they are related to irrelevant operators of dimension $\Delta_f=8$ and $\Delta_w=6$ respectively. The dilaton, instead, has no mass term and as such is dual to a marginal (actually a marginally irrelevant) operator.

\subsubsection*{The action to first order}\label{subfir}
 From the above expressions we see that, to first order in $\epsilon_*$, the 5d effective action reduces to just an Einstein-Hilbert action with negative cosmological constant. This has thus a simple $AdS_5$ (black hole) solution 
\be
ds_5^2 = \frac{r^2}{L^2}\left(1-\frac{\epsilon_*}{3}\right) [- b_0(r) dt^2 + dx_i dx_i] + L^2 \left(1+\frac{\epsilon_*}{3}\right) \frac{dr^2}{r^2 b_0(r)}\,,\quad b_0(r) = 1- \frac{r_h^4}{r^4}\,,
\label{umf}
\ee
with effective radius $L_f$ given by $L_f^2 = L^2 [1+\epsilon_*/3]$, where $L$ is the $AdS$ radius of the unflavored model. The thermodynamics to first order in $\epsilon_*$ can be thus easily obtained. All one needs is simply to replace $L$ with $L_f$ in the $AdS$ black hole thermodynamical formulas. Thus, using $T=r_h/(\pi L_f^2)$ for the temperature, the entropy density $s= [1/(4 G_5)] \pi^3 T^3 L_f^3$ as well as the other thermodynamical observables follow. The results are in perfect agreement with those found in \cite{Bigazzi:2009bk} using the 10d solution\footnote{Notice that the standard $AdS$ radial coordinate used in (\ref{umf}) and the $r$-coordinate used in \cite{Bigazzi:2009bk} are simply related by a rescaling at first order: $r_{\rm{here}}=[1+(5\epsilon_*/24)] r_{\rm{there}}$.} and with the results obtained in the probe approximation \cite{Mateos:2007vn}. 

\subsubsection*{The action to second order}\label{secondunch} 

At second order in $\epsilon_*$ the scalar fields start playing a r\^ole in the effective action. The equations of motion for $f_1$ and $w_1$ (dual to irrelevant operators) admit simple trivial solutions $f_1=w_1=0$. These precisely match with the solutions found in \cite{Bigazzi:2009bk} when all the power-like cutoff-suppressed terms are neglected. In this way these fields are effectively integrated out and the constant values entering in the redefinitions (\ref{redef}) of $f,w$ can thus be seen as the IR remnant of integrating out the irrelevant fields.

The remaining non-trivial part of the effective 5d action, which only contains the metric and the scalar field $\phi_1$, can be rewritten, to second order in $\epsilon_*$, as a simple Einstein-dilaton action of the Chamblin-Reall \cite{Chamblin:1999ya} kind
\be
S_{eff}=\frac{1}{16\pi G_5}\int d^5x \sqrt{-\det g}\left[R[g] - \frac12 (\partial\varphi)^2 - V_0 e^{\gamma\varphi} \right]\,,
\ee
where $\varphi\equiv - \epsilon_* \phi_1$ and, to the order we are interested in,
\be
V_0 = -12 \left[1-\frac{\epsilon_*}{3}-\frac{\epsilon_*^2}{72}\right]\,,\quad \gamma=\frac{\epsilon_*}{3}\,.
\ee
Despite the fact that we have to stop at second order in $\epsilon_*$, let us recall, reporting e.g. the results in \cite{gubserpufu}, that this action has a known exact black hole solution. In the $r=\varphi$ ``gauge'' it reads
\be
ds_5^2 = e^{2A(\varphi)}[-h(\varphi) dt^2 + dx_idx_i] + e^{2B(\varphi)}\frac{d\varphi^2}{h(\varphi)}\,,
\ee
where
\bea
e^{2A}&=& {\rm exp}\left(-\frac23\frac{\varphi}{\gamma}\right)\,,\qquad e^{2B}=-\frac{8-3\gamma^2}{6\gamma^2 V_0}{\rm exp}(-\gamma\varphi)\,,\nonumber\\
h&=& 1- {\rm exp}\left[-\frac{8-3\gamma^2}{6\gamma}(\varphi_h-\varphi)\right]\,.
\eea
From these formulae we get the entropy density and temperature of the black hole as (see also \cite{gubsernello})
\bea
s &=& \frac{e^{3A_h}}{4G_5} = \frac {e^{-\frac{\varphi_h}{\gamma}}}{4 G_5}\,,\nonumber \\
T&=& \frac{e^{A_h-B_h}}{4\pi} |h'(\varphi_h)|= \frac{1}{4\pi}\sqrt{-\frac{4}{3}V_0}\sqrt{1-\frac38\gamma^2}\,e^{-\frac{\varphi_h}{6\gamma}(2-3\gamma^2)}\,,
\eea
so that, to second order,
\be
s = \sigma_{0} T^3 \left[1+\frac{\epsilon_h}{2}+\frac{7}{24}\epsilon_h^2\right]\,,
\ee
where
\be
\epsilon_h = \epsilon_* + \epsilon_*^2 \log (\pi T)\,,
\ee
so that
\be
T\frac{d\epsilon_h}{d T} = \epsilon_h^2\, .
\ee
Above, $\sigma_0 T^3$ is the entropy density of the unflavored plasma, with 
\be
\sigma_0 = \frac{\pi^3L^3}{4 G_5} = \frac{\pi^5 N_c^2}{2 V(X_{SE})}\,,
\label{sigmazero}
\ee
being a measure of the number of degrees of freedom of the unflavored theory (it is proportional to the central charge holographically given by $a= N_c^2 \pi^3 /(4 V(X_{SE})$).
The entropy density, as well as the ``RG-running" formula for $\epsilon_h$ given above, precisely coincide with the ones  found using the full 10d action in \cite{Bigazzi:2009bk}. The remaining thermodynamical observables also  follow accordingly. 

These results confirm that in the deep IR limit (i.e. for $T\ll\Lambda_s\ll\Lambda_{UV}$) the thermodynamics is captured by just the marginally irrelevant operator dual to the dilaton field. The same conclusion holds for the hydrodynamic behavior, as it has been shown in \cite{studies1}.  

\subsection{The charged black hole solution}
Let us consider now the charged case, which is the focus of the present paper. In the following we will adopt the same redefinitions for the scalar fields as in (\ref{redef}) together with the following ones for the forms
\be
{\cal A}_1 = e^{\Phi_*/2} A\,,\qquad d {\cal C}_2= Q_f e^{\Phi_*/2}F_3\,,\qquad {\cal C}_1=Q_f e^{\Phi_*/2} V\,.
\label{reform}
\ee
With these redefinitions we find, at second order in $\epsilon_*$
\be
S_{eff}=\frac{1}{16\pi G_5}\int dx^5 \sqrt{-\det g}\left[R[g]+12-4\epsilon_* \sqrt{1+\frac{F^2}{2}}+\epsilon_*^2 {\cal L}_2\right]\,,
\label{eq.effectivechargedaction}\ee
where $F=dA$, $Y=dV$ and $F_3=dC_2$. The term ${\cal L}_2$ is given by
\bea
{\cal L}_2 &=& -\frac{(\partial f_1)^2}{2} - \frac{(\partial w_1)^2}{2} -\frac{(\partial \phi_1)^2}{2} -\frac56 + \frac23 \Lambda_1[F^2]+\nonumber\\
&& - G_1[F^2] f_1 - H_1[F^2] w_1 -16 f_1^2 - 6w_1^2 - 4\phi_1 \Lambda_1[F^2]+\nonumber\\
&& -\frac12 (Y+F)^2 -\frac12 F_3^2 - 4 V^2\,,
\eea
with the functionals
\bea
\Lambda_1[F^2]&=&\frac{1 + F^2/4}{\sqrt{1+(F^2/2)}}\,,\nonumber\\
G_1[F^2] &=& \frac{3\sqrt{2} F^2 + 16\sqrt{2} -16 \sqrt{2+F^2}}{\sqrt{15}\sqrt{2+F^2}}\,,\nonumber\\
H_1[F^2]&=& \frac{-2\sqrt{2} + 2\sqrt{2+F^2}}{\sqrt{5}}\,.
\eea
In the uncharged case $F=V=F_3=0$ we get $\Lambda_1=1$, $ G_1=H_1=0$ so that the uncharged second order effective action (\ref{seceffun}) is recovered.
\subsubsection*{The action to first order}
From eq. (\ref{eq.effectivechargedaction}) we see that, at first order in $\epsilon_*$, the effective action reduces to a simple Einstein-DBI one, where only the metric and the $U(1)_B$ field $F$ enter.
The thermodynamics of the charged D3-D7 system at first order in $\epsilon_*$  is thus fully captured by that simple action. Using the electric ansatz given in (\ref{metans}), (\ref{formans}) we find that the equation of motion for the vector field is readily solved by
\be
F_{rt}=\frac{e^{A(r)+B(r)} r_d^3}{\sqrt{e^{6A(r)}+r_d^6}}\,,
\ee
where $r_d$ is a dimensionful constant naturally associated with the charge density per flavor.\footnote{Notice that $r_d$ is not a radial position, since it can have a negative value in correspondence to the sign of the charge density per quark.}

Writing the metric as $ds^2 = g_{MN}(r) dx^M dx^N$ it is easy to see that Einstein's equations\footnote{Cfr. e.g. \cite{pal}, eqns. (18)-(20), setting $V=0, Z_1=Z_2=1, \lambda=1, \phi=0, d=4,\Lambda=-6, T_b=4\epsilon_*$.}
\begin{align}
R_{tt} + 4 g_{tt} + \frac23 \epsilon_* \frac{g_{tt}}{g_{xx}^{3/2}}\sqrt{r_d^6 + g_{xx}^3}-2\epsilon_*\frac{g_{tt}g_{xx}^{3/2}}{\sqrt{r_d^6 + g_{xx}^3}} & = 0\,,\nonumber\\
R_{xx} + 4 g_{xx} - \frac43 \epsilon_* \frac{\sqrt{r_d^6 + g_{xx}^3}}{g_{xx}^{1/2}} & = 0\,,\nonumber\\
R_{rr} + 4 g_{rr} + \frac23 \epsilon_* \frac{g_{rr}}{g_{xx}^{3/2}}\sqrt{r_d^6 + g_{xx}^3}-2\epsilon_*\frac{g_{rr}g_{xx}^{3/2}}{\sqrt{r_d^6 + g_{xx}^3}} & = 0\,,
\end{align}
admit the following simple solutions (found imposing regularity at the horizon and UV matching with the uncharged flavored solution in (\ref{umf}))
\begin{align}
g_{tt} &= - r^2 \left(1-\frac{\epsilon_*}{3}\right) b(r)\,,\nonumber \\
g_{xx}&= r^2\left(1-\frac{\epsilon_*}{3}\right)\,,\nonumber \\
g_{rr}&= \frac{\left(1+\frac{\epsilon_*}{3}\right)}{r^2 b(r)}\,,\nonumber 
\end{align}
where
\begin{align}
b(r)&= \left(1+\frac{\epsilon_*}{3}\right)\left(1-\frac{r_h^4}{r^4}\right)+\epsilon_* b_q(r)\,,\nonumber\\
b_q(r)&= -\frac{1}{3r^3}\sqrt{r^6+r_d^6}+\frac{r_h}{3r^4}\sqrt{r_h^6+r_d^6}-\frac{r_d^4}{r^4}\frac{G_F(r) - G_F(r_h) }{2} \,,\nonumber\\
G_F(r)&= \frac{1}{3} B \left( \frac{r^6}{r^6+r_d^6} ; \frac{1}{6} , \frac{1}{3} \right) = \frac{1}{3^{1/4}} F\left[\cos ^{-1}\left(\frac{\left(1-\sqrt{3}\right)   r^2+r_d^2}{\left(1+\sqrt{3}\right)   r^2+r_d^2}\right) \Bigg|\frac{ 2+\sqrt{3} }{4}\right]  \,.
\label{firstcha}
   \end{align}
$F[x|y]$ is the Elliptic integral of the first kind, $B(x;a,b)$ the incomplete beta function and we have chosen the integration constants so that $b(r_h)=0$. Finally, the electric field reads\footnote{In order to get the first order correction to this expression we need to consider the second order effective action. See below.}
 \be
F_{rt}=\frac{r_d^3}{\sqrt{r_d^6+r^6}}\,. 
\ee 
This result is in agreement with the solution obtained in the probe approximation in \cite{Karch:2007br} (see their eq. (3.8)). 
Integrating the expression above and imposing $A_t(r_h)=0$ we find
\be
A_t(r) = r_d\frac{G_F(r) - G_F(r_h) }{2} \,.
\label{atsol}
\ee
\subsubsection*{The action to second order}
Let us now consider the effective action to second order. If one is interested in the thermodynamics (obtained from the free energy, which in turn is holographically related to the on-shell gravity action), it is clear that the relevant equations of motion for $f_1$, $w_1$, $C_2$, $\phi_1$ and $V$ which appear in the action at ${\cal O}(\epsilon_*^2)$, can be solved just by computing them on the ``zeroth-order'' background
\bea
&& ds_0 ^2 =r^2 [- b_0(r) dt^2 + dx_i dx_i] + \frac{dr^2}{r^2 b_0(r)}\,,\quad b_0(r)= 1- \frac{r_h^4}{r^4}\,, \nonumber \\
&&(F_{rt})_0= \frac{r_d^3}{\sqrt{r_d^6+r^6}}\,,\,\,{\rm so\, that}\,\, (F^2)_0=- 2 \frac{r_d^6}{r_d^6 + r^6}\,.
\eea 

The equations for $f_1$, $w_1$, $\phi_1$, $V$ and $C_2$ which we need to solve are
\bea
(5r^4-r_h^4)f_1' + r (r^4 - r_h^4) f_1'' &=& r^3 (G_1[(F^2)_0] + 32 f_1)\,,\nonumber\\
(5r^4-r_h^4)w_1' + r (r^4 - r_h^4) w_1'' &=& r^3 (H_1[(F^2)_0] + 12 w_1)\,, \nonumber\\
(5r^4-r_h^4)\phi_1' + r (r^4 - r_h^4) \phi_1''&=& 4 r^3 \Lambda_1[F^2]\,, \nonumber \\ 
d\star_0 F_3&=&0\,,\nonumber\\
d\star_0 (Y+F_0) &=& 8 \star_0 V\,, 
\label{eomsec}
\eea
where the Hodge dual is computed on the zero-th order background. It $F=0$ these equations consistently reduce to the uncharged ones, which have $f_1=w_1=0$ as solutions. In the charged case these trivial solutions are not admitted and all the scalar fields are running. As in the last equation $F=F_0$, we see that the field $V$ (a massive vector field dual to an irrelevant operator) is ``effectively'' decoupled from the other fluctuations in this setting. 

The above system of equations is actually not complete: there is in fact a further constraint which should be imposed by hand which comes from the requirement that the 5d $H_3$ fields (which do not appear in our effective 5d action) can be consistently set to zero. In particular, from eq. (78) in \cite{Cotrone:2012um} we get that to leading order
\be
F_3 (r)= r^3 \frac{\sqrt{-(F^2)_0/2}}{\sqrt{1+(F^2)_0/2}} dx^1\wedge dx^2 \wedge dx^3 = r_d^3\, dx^1\wedge dx^2\wedge dx^3\,,
\ee
which readily solves the corresponding equation in \eqref{eomsec}.

The solutions to the rest of the equations (\ref{eomsec}) (found by requiring regularity at the horizon and UV matching\footnote{We present the solution at finite UV cutoff $r_s$.} with the $T=0$ uncharged flavored solution of section \ref{sec:unch}) are quite involved and we can give them just in a semi-analytic form
\begin{align}
\sqrt{\frac{3}{80}} f_1 = & \frac{1}{40} - \frac{1}{90} \frac{2r^4-r_h^4}{2r_s^4-r_h^4}  - (2r^4-r_h^4) \int_r^{r_s} \frac{8\sqrt{\tilde r^6+r_d^6 } }{40(2 \tilde r^4-r_h^4)^2} d \tilde r  \\
\nonumber & \quad + r_h^4 r_d^4 (2r^4-r_h^4) \int_r^{r_s} \frac{   G_F(\tilde r) - G_F(r_h) }{20\tilde r (\tilde r^4 - r_h^4)(2 \tilde r^4-r_h^4)^2} d \tilde r  \ , \\
\sqrt{\frac{1}{40}} w_1 = & \frac{1}{60} +  \frac{P_{1/2} \left(2\frac{r^4}{r_h^4}-1\right)}{P_{1/2} \left(2\frac{r_s^4}{r_h^4}-1\right)} \left[ -\frac{1}{60} - \frac{P_{1/2} \left(2\frac{r_s^4}{r_h^4}-1\right)}{10r_h^4} \int_r^{r_s}\frac{\tilde r^6 R_{1/2} \left(2\frac{\tilde r^4}{r_h^4}-1\right)}{\sqrt{\tilde r^6+r_d^6}} d\tilde r \right] \\
& -\frac{ r^6 R_{1/2} \left(2\frac{ r_s^4}{r_h^4}-1\right)}{10r_h^4} \Bigg[ \frac{R_{1/2} \left(2\frac{r^4}{r_h^4}-1\right)}{R_{1/2} \left(2\frac{r_s^4}{r_h^4}-1\right)} \int^r_{r_h}\frac{\tilde r^6 P_{1/2} \left(2\frac{\tilde r^4}{r_h^4}-1\right)}{\sqrt{\tilde r^6+r_d^6}} d\tilde r  \nonumber\\
& \qquad\qquad\qquad\qquad\quad - \frac{P_{1/2} \left(2\frac{r^4}{r_h^4}-1\right)}{P_{1/2} \left(2\frac{r_s^4}{r_h^4}-1\right)} \int_r^{r_s}\frac{\tilde r^6 P_{1/2} \left(2\frac{\tilde r^4}{r_h^4}-1\right)}{\sqrt{\tilde r^6+r_d^6}} d\tilde r  \Bigg] \nonumber \ ,
\end{align}
where we have used the explicitly real combination $R_{1/2}(x) = Q_{1/2}(x)+\frac{i\pi}{2} P_{1/2}(x)$, with $P_n(x)$ and $Q_n(x)$ the Legendre functions of the first and second kind.

The dilaton solution reads
\begin{align}\nonumber
\phi_1  = &  \left( \frac{r \sqrt{r^6+r_d^6} - r_h \sqrt{r_h^6+r_d^6} }{4r_h^4}\right) \log b_0(r) - \left( \frac{r_* \sqrt{r_*^6+r_d^6} - r_h \sqrt{r_h^6+r_d^6} }{4r_h^4}\right) \log b_0(r_*) \\
& +  \frac{r_d^4}{8r_h^4} \left( D_F(r)- D_F(r_*)  \right) + \frac{1}{2r_h^4} \int_r^{r_*} \frac{2 \tilde r^6 + r_d^6}{\sqrt{\tilde r^6 + r_d^6}} \log b_0(\tilde r) d\tilde r \ .  \label{eq.phi1solution}
\end{align}
where we have  defined
\be
D_F(r)  = \log  b_0(r)  \left( G_{F}(r) - G_F(r_h) \right) \ ,
\ee
and finally, for the vector field, we write
\be\label{eq.Vtredef}
V_t = 4 r_d^3 {\cal J}_1 ' \frac{r^4-r_h^4}{r^5}\,,
\ee
so that the equation of motion can be written as\footnote{Recall that the parameter $\sigma$ is equal to $-1$ in our setup and to $1$ in the anti-D7 brane case.}
\be
\partial_r \left( \frac{r^4-r_h^4}{r^5} \cJ_1'  \right)  - 8 r^{-3}  \cJ_1  =  \frac{\sigma }{4 \sqrt{r^6+r_d^6}}  \ ,
\ee
with solution
\begin{align}
\cJ_1 =&  \sigma  \left(2 \frac{ r^2}{r_h^2}+\left(  1 + \frac{r^4}{r_h^4}\right)  \log \left[\frac{r^2-r_h^2}{r^2+r_h^2}\right]\right)  \Bigg[ \frac{\left(\left(\sqrt{3}-1\right  ) r_d^4+2 r_h^4\right)}{256 r_d^2  r_h^2}  (G_F(r)-G_F(r_h)) \\
&  + \frac{1}{128 r_h^2} \left(\frac{ \left(1+\sqrt{3}\right) r \sqrt{r^6+r_d^6}}{  \left(\left(1+\sqrt{3}\right) r^2+r_d^2\right)}  - \frac{  \left(1+\sqrt{3}\right) r_h\sqrt{r_d^6+r_h^6}}{\left(r_d^2+\left(1+\sqrt{3}\right) r_h^2\right)}  \right) \nonumber \\ 
& -\frac{ \sqrt{3} r_d^2}{128 r_h^2} (G_E(r)-G_E(r_h)) \Bigg]  - \sigma \frac{r^4+r_h^4}{r_s^4+r_h^4} \left( 2 \frac{  r_s^2}{r_h^2}+\left(1 + \frac{r_s^4}{r_h^4}\right) \log  \left[\frac{r_s^2-r_h^2}{r_h^2+r_s^2}\right] \right)\times \nonumber \\   
& \Bigg[  \frac{\left(\left(\sqrt{3}  -1\right) r_d^4+2 r_h^4\right)  }{256 r_d^2 r_h^2} (G_F(r_s)-G_F(r_h)) \nonumber\\
&  +\frac{1}{128 r_h^2} \left( \frac{ \left(1+\sqrt{3}\right) r_s \sqrt{r_d^6+r_s^6}}{  \left(r_d^2+\left(1+\sqrt{3}\right) r_s^2\right)}-\frac{ \left(1+\sqrt{3}\right) r_h \sqrt{r_d^6+r_h^6}}{  \left(r_d^2+\left(1+\sqrt{3}\right)  r_h^2\right)} \right)  \nonumber\\
&  -\frac{ \sqrt{3} r_d^2  }{128 r_h^2}  (G_E(r_s)-G_E(r_h)) \Bigg]
 + \sigma \frac{r^4+r_h^4}{64 r_h^6} \int_r^{r_s}  \frac{\left(\tilde r^4+r_h^4\right) }{\sqrt{\tilde r^6+r_d^6}}\log  \left[\frac{\tilde r^2-r_h^2}{\tilde r^2+r_h^2}\right] \,  d\tilde r  \nonumber\\
& - \frac{\sigma}{96}  \left(1+ \frac{r^4}{r_h^4}\right) \log  \left[\frac{r^3 + \sqrt{r^6+r_d^6}}{r_s^3 + \sqrt{r_d^6+r_s^6}}\right]    \ , \nonumber
\end{align}
where $G_E$ is defined in the same way as $G_F$ in \eqref{firstcha}, but with the elliptic function of the second kind $E[x|y]$ instead of the first kind one $F[x|y]$.
 In particular, notice the linear dependence with the WZ factor $\sigma$, describing the dependence of this function on the charge of the smeared branes.

While for the other fields the zero-th order background suffices, in order to solve for $F_{rt}$ we need the first order background metric (\ref{firstcha}), which has $g_{tt} g_{rr} =-1$ and $\sqrt{g} = r^3[1-(\epsilon_*/2)]$. Dubbing $R_d$  the integration constant, we find that, to first order
\be\label{eq.Frtcorrection}
F_{rt}(r) = \frac{R_d^3}{\sqrt{r^6 + R_d^6}} \left[1+ \epsilon_* \frac{r^6}{r^6+R_d^6} Q[r]\right]\,,
\ee
where $Q[r]$ is expressed in terms of the on-shell values of the scalar fields and the vector $V$
\be
Q[r]=\frac12\left[1-\frac{r^3}{2 R_d^3}\frac{\partial {\cal L}_2}{\partial F_{rt}}|_0\right]\,,
\ee
and the derivative has to be evaluated on the zero-th order solution for $F_{rt}$.

The integration constant $R_d$ can be written as an expansion
\be
R_d= r_d \left(1+\frac{\epsilon_*}{3} \rho_1 \right)\,,
\ee
with $\rho_1\ll 3/\epsilon_*$. This way we can equivalently write
\be\label{eq.Frtcorrectionbis}
F_{rt}(r) = \frac{r_d^3}{\sqrt{r^6 + r_d^6}} \left[1+ \epsilon_* \frac{r^6}{r^6+r_d^6} (\rho_1+ Q[r])\right]\,,
\ee
with (to be explicit)
\be
Q[r]=\frac12\left[1+\frac{(-1+6\phi_1)r_d^6}{6r^6}+\frac{f_1(2r^6+5r_d^6)}{\sqrt{15}r^6}-\sqrt{\frac{2}{5}}w_1-r^3\left(\frac{1}{\sqrt{r^6+r_d^6}}+\frac{Y_{rt}}{r_d^3}\right)\right]\,.
\ee
\subsubsection*{Comments}
The whole charged solution described in this section is perturbative in $\epsilon_*$ but exact in $r_d$. In \cite{Bigazzi:2011it} the solution with $|r_d|\ll r_h$ was presented, which implied a second perturbative expansion in $\delta \equiv r_d^3 / r_h^3 $. 
The solution in \cite{Bigazzi:2011it} is thus only valid when the energy scale of the charge density per flavor (or chemical potential if a change of ensemble is performed) is much lower than the energy scale associated to the temperature. The fact that our solution is exact in $r_d$ does not mean that the solution is valid for asymptotically large values of the charge density (or chemical potential) per flavor, since the solution is still perturbative in $\epsilon_*$ and large values of $r_d$ may push the solution out of the region of validity. This will be clarified in the following sections by studying the extremal limit. This limit requires the knowledge of  the solution at second order in $\epsilon_*$. This solution is more involved and we have not been able to find a complete analytic expression for the metric in that case (the explicit solutions to $f_1$, $w_1$, $\phi_1$, etc. enter in the equations of motion, implying that an analytic approach becomes an extremely convoluted task), however we can obtain crucial informations that allow us to find some conclusive results. 

%%%%%%%%%%%%%%%%%%%%%%%%%%%%%%%%%%%%%%%%%%%%%%%%%%%%
\section{Thermodynamics}\label{sec.thermo}

In this section we discuss the thermodynamics of the charged solution \emph{at first order} in $\epsilon_*\sim\epsilon_h$, where we have an analytic solution for the 5d metric and the electric $U(1)_B$ field.\footnote{Extrapolations to QGP RHIC temperatures give rise to $\epsilon_h\sim0.24$ \cite{Bigazzi:2009bk} and we will use this value where needed in our plots.} 

To begin with, let us notice that the thermodynamics of the same system has been calculated \emph{in the probe approximation} in \cite{Kobayashi:2006sb}.
Let us immediately make it clear that, in the cases we can compare, the thermodynamical quantities calculated here and in  \cite{Kobayashi:2006sb} will coincide \emph{modulo the temperature}.
That is, the observables will coincide once we write the result in \cite{Kobayashi:2006sb}, where the temperature was the constant % (in the charge density per quark) 
temperature of the unflavored theory, in terms of the temperature of the backreacted solution.\footnote{This was the case also in the perturbative solution in \cite{Bigazzi:2011it}.}

The temperature of the charged black hole solution found in the previous section reads
 \be
 T=\frac{r_h}{\pi}\left[1-\frac{\epsilon_h}{3}\sqrt{1+\frac{r_d^6}{r_h^6}}\right]\equiv T_0 \left[1-\frac{\epsilon_h}{3}\sqrt{1+\frac{r_d^6}{r_h^6}}\right]\,,
\label{temperature}
\ee
where  $T_0$ is the temperature in the unflavored case. Equation (\ref{temperature}) is a monotonic function of $r_d$ (namely, of the chemical potential or charge density per flavor) vanishing at 
\be
r^{extr}_d = \left[\frac{9}{\epsilon_h^2} -1\right]^{1/6} r_h\,.
\label{extcon}
\ee
We do not expect, however, that we can really reach the extremality regime where $T \sim 0$ since there the corrections to $T_0$ are order unity and we are not guaranteed that higher orders in $\epsilon_h$ will be subleading. In fact, we will argue in section \ref{sec.extremal} that the corrections are not subleading, so that the extremal perturbative solution is not reliable.

The field theory entropy density, $s$, holographically identified with the Bekenstein-Hawking entropy of the dual black hole
\be
s=\frac{1}{4G_5} r_h^3 \left(1-\frac{\epsilon_h}{2}\right)\,,
\ee
can be recast in terms of the temperature as
\be
s=\sigma_0 T^3 \left[1-\frac{\epsilon_h}{2}+\epsilon_h\sqrt{1+\delta^2}\right]\,.
\ee
where 
\be
\delta=\frac{r_d^3}{r_h^3}\,, 
\ee
and $\sigma_0$ defined as in $(\ref{sigmazero})$.
 It is easy to see that the entropy density precisely reduces to that found in \cite{Bigazzi:2011it} in the $\delta\ll1$ limit at second order in $\delta$.
 
The charge density $n_q$ is given by the holographic relation
\be
n_q = \frac{(2\pi\alpha')}{L^5} e^{-\Phi_*/2}\frac{d S}{d \dot A_t}=(2\pi\alpha') e^{-\Phi_*/2}\epsilon_h\frac{r_d^3}{4\pi G_5 L^5}\,,
\label{nq}
\ee
so that
\be
\frac{n_q}{T^3} = \epsilon_h \sqrt{\frac{\pi^7}{\lambda_h V(X_{SE})}} N_c^2 \delta = \frac{V(X_3)}{16\pi}\frac{\pi^{7/2}}{V(X_{SE})^{3/2}}\sqrt{\lambda_h} N_f N_c\, \delta \,.
\label{dnq}
\ee
This relation matches (to first order in $\epsilon_h$) with that obtained in \cite{Bigazzi:2011it} in the small $\delta$ limit. This is related to the fact that (no matter the value of $\delta$) at this order the field $A_t$ is effectively decoupled from the vector and two-form fields in the action (\ref{5dcharged}). Notice that the baryon charge density $n_B$ is related to $n_q$ by the simple relation $n_q = N_c n_B$.

From (\ref{dnq}) we get that at first order in $\epsilon_h$ and fixed charge density (canonical ensemble)
\be
\left(\frac{d\delta}{d T}\right)_{n_q} = -\frac{3\delta}{T}\,.
\ee
Using this result it is easy to verify that the Helmholtz free energy density, given in terms of the incomplete beta function
\bea
f &=& - \frac14 \sigma_{0} T^4 \left[1-\frac{\epsilon_h}{2}\left(1- 2\sqrt{1+\delta^2} + \Sigma(\delta)\right)\right]\,,\nonumber \\
\Sigma(\delta)&=& \delta^{4/3} \left[ B\left(1;\frac{1}{6},\frac{1}{3} \right)-B\left(\frac{1}{1+\delta^2};\frac{1}{6},\frac{1}{3} \right) \right] \,,
\eea
satisfies the thermodynamical relation $s=-\partial f/\partial T$ (canonical ensemble). Notice that $\Sigma(\delta)\approx 3 \delta^2$ in the $\delta\rightarrow0$ limit. Using this result we see that the formula above precisely reduces to the corresponding one found in \cite{Bigazzi:2011it} in the small $\delta$ limit.

The energy density (holographically related to the ADM black hole one) reads
\be
e=\frac34\sigma_{0} T^4 \left[1-\frac{\epsilon_h}{2}\left(1- 2\sqrt{1+\delta^2} - \frac13\Sigma(\delta)\right)\right]\,,
\ee
and can be deduced from the relation $e=f+s T$.  From this expression we easily determine the pressure $p$, and thus the Gibbs free energy density $\omega$, as
\be
p=-\omega=\frac{e}{3} + {\cal O}(\epsilon_h^2)\,.
\ee
This also follows from the fact that at first order in $\epsilon_h$ the trace of the stress energy tensor at thermodynamical equilibrium does not receive corrections with respect to its (zero) value in the conformal limit. In fact we know that the conformality breaking effects are  higher orders in $\epsilon_h$ \cite{Bigazzi:2009bk} and that  the chemical potential does not contribute to the conformality breaking in a theory like the present one where matter fields are massless. This also implies that the speed of sound is $v_s^2 = 1/3$ and the specific heat at fixed chemical potential is $c_{|\mu}=3 s$.

\begin{figure}[tb]
\begin{center}
\includegraphics[scale=0.7]{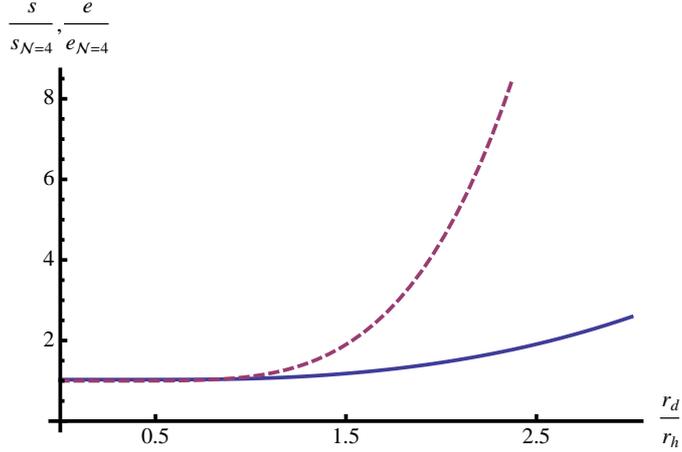}
\caption{ $s/s_{{\cal N}=4}$ (solid line) and $e/e_{{\cal N}=4}$ (dashed line) as functions of $r_d/r_h$ with $\epsilon_h=0.24$.}
\label{fig.all}
\end{center}
\end{figure}

In figure \ref{fig.all} we focus on the flavored ${\cal N}=4$ SYM case and we report the entropy density and the energy density, normalized with respect to the ones in the unflavored theory. 
Nothing special happens at the would-be ``extremality point'' $r_d^{extr} \sim 2.32 r_h$.
In particular, the entropy density is finite.

As said above, in order to compare with the probe results in \cite{Kobayashi:2006sb}, we have to normalize our quantities by the appropriate power of $T/T_{{\cal N}=4}$.
After this normalization, the agreement of our results with the ones in \cite{Kobayashi:2006sb} is perfect.

\begin{figure}[tb]
\begin{center}
\includegraphics[scale=0.7]{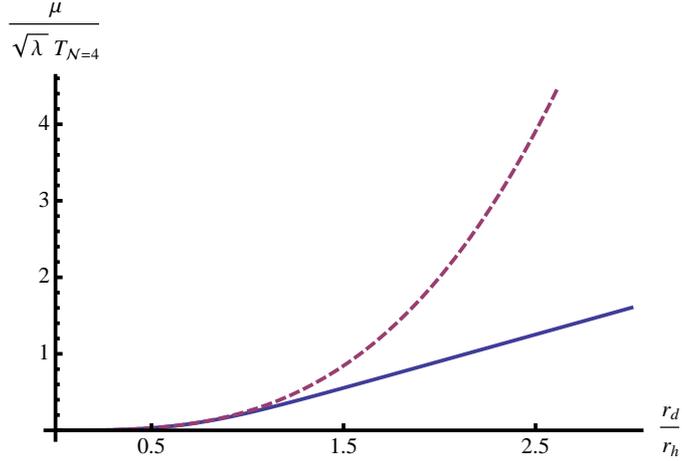}
\caption{ $\mu/\sqrt{\lambda}T_{{\cal N}=4}$ as a function of $r_d/r_h$ in the perturbative charged (dashed line) and exact charged (solid line) cases, with $\epsilon_h=0.24$.}
\label{fig.mu.compare}
\end{center}
\end{figure}

We can check that the thermodynamical relation $s=-\partial\omega/\partial T$ at fixed chemical potential (gran-canonical ensemble) is consistently satisfied . First of all notice that from the thermodynamical relation $\mu n_q = f-\omega$ and from formulas (\ref{nq}), (\ref{dnq})
we get (to leading order)
\be
\frac{\mu}{T} = \frac{\pi}{12} \frac{\Sigma(\delta)}{\delta}\sqrt{\frac{\pi\lambda_h}{V(X_{SE})}}\,.
\label{chempot}
\ee
It can be checked that, consistently,  
\be
\mu=\frac{e^{\Phi_*/2}}{2\pi\alpha'}A_t(\infty) \,,
\ee 
where $A_t(r)$ is given in (\ref{atsol}). In the small $\delta$ limit these expressions precisely agree (to first order in $\epsilon_h$) with those given in \cite{Bigazzi:2011it}. In particular in this limit we have $\mu/T\sim\delta$ hence $\mu/T$ scales like $r_d^3$. In the opposite $\delta\gg1$ limit (which can be formally taken with the proviso of taking $r_d$ not greater than $r_d^{ext}$) we get
\be
\frac{\mu}{T}\approx \frac{\pi}{12}\sqrt{\frac{\pi\lambda_h}{V(X_{SE})}} \left[ B \left( 1 ; \frac{1}{6} , \frac{1}{3} \right)  \delta^{1/3} - 6 \right] \sim  \frac{\pi}{12}\sqrt{\frac{\pi\lambda_h}{V(X_{SE})}} \left[8.4 \frac{r_d}{r_h} - 6 \right]\,, \quad (\delta\gg1)\,,
\ee
which shows that $\mu/T$ scales linearly with $r_d$. This behavior is made evident in figure \ref{fig.mu.compare} where (focusing on the ${\cal N}=4$ SYM case) we also compare our results against the perturbative-in-$\delta$ case, normalized with $\sqrt{\lambda_h}\, T_{{\cal N}=4}$. 
From (\ref{chempot}) we get that 
\be
\left(\frac{d\delta}{dT}\right)_{\mu} = - \frac{3\delta}{T}\left(\frac{\sqrt{1+\delta^2}\,\Sigma(\delta)}{6\delta^2+\sqrt{1+\delta^2}\,\Sigma(\delta)}\right)\,,\ee
which is what we need to verify that $s=-\partial\omega/\partial T$ at fixed $\mu$.

\begin{figure}[tb]
\begin{center}
\includegraphics[scale=0.7]{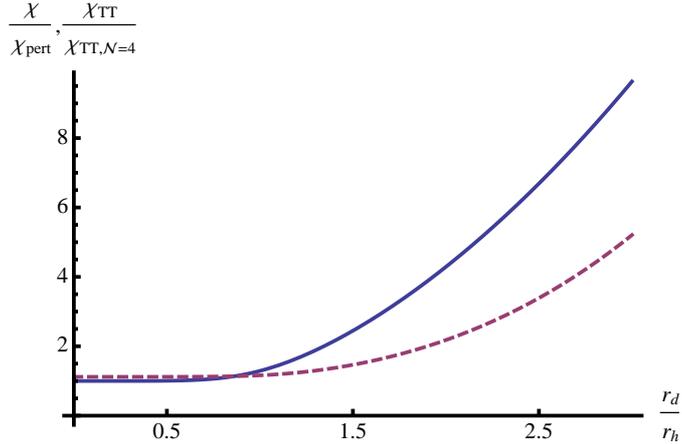}
\caption{ $\chi/\chi_{pert}$ (solid line) and $\chi_{TT}/\chi_{TT,{\cal N}=4}$ (dashed line) as functions of $r_d/r_h$ with $\epsilon_h=0.24$.}
\label{fig.chi.all}
\end{center}
\end{figure}

Now we are in the position to calculate the susceptibility matrix, in order to check the thermodynamical stability of the system. The ``quark'' susceptibility $\chi=\partial_{\mu} n_q$ and $\chi_{TT}= - \frac{\partial^2 \omega}{\partial^2 T}$ are reported in figure \ref{fig.chi.all} for the flavored ${\cal N}=4$ SYM master example.
As we can see, both susceptibilities are increasing positive functions of the chemical potential.
Obviously, their values at $\mu=0$ are the same as in the perturbative case
\begin{equation}
\chi_{pert}=\frac{4\pi^2}{\lambda} \epsilon_h N_c^2 T^2\ , \qquad \chi_{TT,{\cal N}=4}=\frac{3\pi^2}{2} N_c^2 T^2\ . 
\end{equation}

The off-diagonal susceptibilities are as in the perturbative case 
\begin{equation}
\chi_{T\mu}=\chi_{\mu T}= \partial_T n_q= \frac{2\pi^2}{\sqrt{\lambda}} \epsilon_h N_c^2 \left( \frac{r_d}{r_h} \right)^3 T^2 \ .
\end{equation}
Actually, we do not need their expression at leading order in $\epsilon_h$ for the issue of stability (i.e. the positivity of the determinant of the susceptibility matrix).
The point is that these observables are already of order $\epsilon_h$, so they are subleading in the determinant, which at leading order results to be simply proportional to the ``quark'' susceptibility $\chi$ (times the unflavored part of $\chi_{TT}$, i.e. $\chi_{TT,{\cal N}=4}$).

In conclusion, the thermodynamical stability of the system is guaranteed by the positivity of the ``quark'' susceptibility $\chi$ for every charge value.

%%%%%%%%%%%%%%%%%%%%%%%%%%%%%%%%%%%%%%%%%%%%%%%%%%%%%
%%%%%%%%%%%%%%%%%%%%%%%%%%%%%%%%%%%%%%%%%%%%%%%%%%%%%
\section{The issue of extremality}\label{sec.extremal}

As we will see in the following, the charged solution in section \ref{sec.chargedblackbrane} is not complete around $T \sim 0$, forbidding us to trust it at extremality.
The point is that going to $T=0$ would require, strictly speaking, going beyond the perturbative-in-$\epsilon$ regime. In fact, at first order, we have found that $T=T_{0}[1- {\cal O}(\epsilon_h)]$ and implementing the extremality condition (\ref{extcon}) would imply taking the subleading ${\cal O}(\epsilon_h)$ terms to be of order one. The perturbative approach would have some hope to be sensible only if, despite being at extremality, the second order terms keep staying subleading with respect to the first order ones. As we will show in the following, this, not unexpectedly, is not the case. 

The extremality condition for the charged solution at first order can be rewritten as
\be\label{tzerof}
y = \frac{\epsilon_h}{3\sqrt{1-\frac{\epsilon_h^2}{9}}} = \frac{\epsilon_h}{3} + {\cal O}(\epsilon_h^3)\,.
\ee
where $y\equiv\delta^{-1}=r_h^3/r_d^3$. In this way we can say that turning on the flavor backreaction, the extremality condition $y=0$ turns into $y= \epsilon_h/3$ at first order in $\epsilon_h$. 

It is easy to show that taking this ``first order extremality limit" the thermodynamic relations found in the previous section precisely reduce to those found at $T=0$ in the charged, massless, D7 probe approximation \cite{Karch:2007br}. In particular the extremal metric develops an $AdS_2$ near-horizon region and the entropy density $s\sim r_d^3$ is non-zero. The free energy scales like $r_d^4$ accordingly.

Now, in view of the previous observations, the question is whether this first order extremality limit makes sense. To answer this question we may ask how the extremality condition is modified at second order. For dimensionality reasons (there are no other scales in the deep IR apart from $r_h$ and $r_d$) the $T=0$ condition to second order should read as
\be
y - \frac{\epsilon_h}{3}\sqrt{1+y^2} - \epsilon_h^2 C[y]=0\,.
\label{extremg}
\ee
Now, if $C[y]$ is an analytic function of $y$, we can expand it around zero getting
\be
y = \frac{\epsilon_h}{3} + \epsilon_h^2 C[0] + {\cal O}(\epsilon_h^3)\,.
\ee
In this case, the second order corrections would be subleading and the first order extremality limit would not be spoiled.  

If instead, say, $C[y]\approx k y^{-m}$ with $m>0$ when $y\rightarrow0$ the situation would be very different.  Take, for example, $m=1$. In this case the first order value $y=\epsilon_h/3$ would be modified to $y= (\epsilon_h/6) (1+\sqrt{1+36k})$. This would mean that the perturbative expansion in $\epsilon_h$ would not be reliable when going to $T=0$: higher order corrections to the $T=0$ condition would modify the critical value of $y$ to lower orders. For $m=2$ we would get that the critical value of $y$ gets a zero-th order correction. This, again, would tell us that the perturbative expansion in $\epsilon_h$ is not reliable when going to $T=0$.

In the following, examining some properties of our charged solution to second order, we will find that, unfortunately, the above situation (actually the $m=1$ case) is precisely realized.
 
 For our purposes, since we are just interested in the $T=0$ condition, it suffices to consider the near-horizon behavior of the matter fields in the small $y$ limit, with $y\equiv r_h^3/r_d^3$. 
From the analytic expressions reported in section \ref{sec.chargedblackbrane}, we find that 
\be
V_t(r_h)=0\,, \quad Y_{rt}(r_h)=V_t'(r_h)=\frac{3}{5}(\pi+\log4-1) + {\cal O}(y^2) \sim 2.12 + {\cal O}(y^2)\,,
\ee
\be
f_1(r_h) = - \frac{4+3\pi-18 \log(2)}{4\sqrt{15}} \frac{1}{y} + ... \sim -\frac{0.06}{y} + ...
\ee
and
\be
w_1(r_h) = \frac{1}{3\sqrt{10}} + ... \sim 0.1 + ...
\ee

From these we see that, among these modes, the leading contribution to the IR physics in the extremality regime comes from the field $f_1$. The horizon contribution of $w_1$, as well as that of the field strength $Y$ of the massive vector mode $V$, is a constant. It would be interesting to precisely understand why a different role is played by these fields, despite the fact that they are all dual to irrelevant deformations. 

As for the dilaton, its value at the horizon can be reabsorbed by a redefinition of $\epsilon$. In fact, as we know,
$\epsilon_h = Q_f e^{\Phi(r_h)} = \epsilon_* (1+ \epsilon_* \phi_1(r_h))$.
%\bea

The horizon value of $F_{rt}$ from \eqref{eq.Frtcorrectionbis} turns out to be
\be
F_{rt}(r_h)= \frac{1}{\sqrt{y^2+1}}\left[1+\epsilon_* \frac{y^2}{y^2+1} \left(\rho_1+ Q[r_h]\right)\right]\,,
\ee
and in the $y\rightarrow0$ limit\be
Q[r_h]= \frac{18\log2 -4 -3\pi}{24 y^3} + {\cal O}(1/y^2) \equiv \frac{a}{y^3}+ {\cal O}(1/y^2)\,,\quad a\sim -0.04\,,
\ee
were we have used the fact that at first order $\epsilon_*\phi_1(r_h)\approx0$. From this expression we see that the $T=0$ extremality condition $y\sim \epsilon_*/3$ which emerges at first order cannot be consistent with the perturbative expansion to second order. In fact, if $y\sim\epsilon_*/3$ the ${\cal O}(\epsilon_*)$ correction in the expression for $F^2(r_h)$ turns out to be not parametrically suppressed: it is just a number equal to $3a\sim -0.12$.\footnote{Recall that by construction $\rho_1\ll 3/\epsilon_*\sim1/y$.} This should be sufficient to state that the second order correction to the $T=0$ condition cannot be made parametrically subleading with respect to the first order one.

To strengthen the above statement let us proceed as follows. Defining the off-shell matter Lagrangian as
\be
L_m = -4\epsilon_*\sqrt{1+F^2/2} +\epsilon_*^2 {\cal L}_2\,,
\ee
we know that Einstein's equation are 
\be
R_{MN} -\frac{R}{2}g_{MN} -6 g_{MN} = \frac12 g_{MN} L_m - \frac{\partial L_m}{\partial g^{MN}}\,.
\ee
In order to get, from these, informations on the $T=0$ condition, it suffices to focus on the $tt$ component. Our metric ansatz will be of the generic form written in (\ref{metans}) and, as usual, we will require that $b(r_h)=0$, so that $b(r)= b'(r_h) (r-r_h) + {\cal O}((r-r_h)^2)$ close to the horizon.

The $T=0$ condition is realized when 
\be
b'(r_h)=0\,.
\ee
Now, taking the $tt$ component of Einstein's equation, multiplied overall by $g^{tt}$, expanding around the horizon, taking just the leading zero-th order terms (in an expansion in $r-r_h$), and using the on-shell horizon values of the various matter fields, we get that the $T=0$ condition is, as expected, precisely of the form written in  (\ref{extremg}).

Moreover, we also discover that, in the $y\rightarrow0$ limit, the function $C[y]$ goes as
\be
C[y] = \frac{(4+3\pi-18\log2)(1+2\pi-12\log2)}{120y} + ...\equiv \frac{k}{y}+... \,,
\ee
so that we fit in the $m=1$ case discussed above.%\be
%y = \frac{\epsilon_h}{3} + \epsilon_h^2 \frac{k}{y}+...\,,
%\ee

These results confirm that in order to reach the $T=0$ regime in the charged D3-D7 models with massless flavors we need to abandon the perturbative-in-$\epsilon$ approach. This implies in particular that {\it the extremality limit cannot be taken in the probe approximation}. This is something which has already been stated in \cite{Hartnoll:2009ns} but perhaps not fully appreciated in the literature. The reason for the breakdown of the reliability of the probe approximation is simple: if $T=0$ the energy density of the unflavored conformal theory (say, ${\cal N}=4$ SYM) is zero. We cannot thus work in the probe approximation consistently, since even a very small number of charged flavors would provide a larger contribution to the energy density. For the same reason, in the case of theories displaying a dynamical scale $\Lambda_{IR}$, the reliability of the probe approximation could be obtained only for a limited range of values of $\mu/\Lambda_{IR}$. 

A final comment is in order. The DBI action we have been using constitutes an approximation to the 
full dynamical flavor effects. In particular, it re-sums the ``one-window graphs'' in the Veneziano 
limit, that is the graphs with one quark loop. At order $\epsilon^2$ one generically expects further corrections, e.g. two-window graphs, to be present. Moreover, despite the fact that they are unknown in the D7-brane case, there could be ``thermal'' corrections (given by the blackfold 
approach \cite{blackfold}) taking into account the proper thermalization of the brane degrees of freedom. 
In order for these corrections to spoil the result of this section, they should all conspire to cancel the non-subleading $\epsilon^2$ contribution discussed above. While we cannot discard this possibility, we find it very unlikely (for example, we suspect that in the $T\ll \Lambda_{UV}$ regime we work in, thermal corrections would be suppressed); were it realized, the solutions presented in section \ref{sec.chargedblackbrane} would include bona-fide extremal charged black branes, with $AdS_2 \times \mathrm{R}^3$ near horizon geometry.

%
%%
%%%
%%%%	NEW SECTION
%%%
%%
%

\section{Perturbations} \label{sec.perturbations}

We  consider now linear fluctuations of the 5d fields around the charged solution. 
We refer again the reader to \cite{Cotrone:2012um} for a complete list of the equations of motion of the 5d system, from where the equations of motion of the fluctuations can be derived by linearizing around the setup described in section \ref{sec.chargedblackbrane}.

Representing generic fluctuations of type IIB supergravity fields by $\Lambda$ and  fluctuations of D7 worldvolume fields by $\lambda$, we can expand   in powers of the backreaction parameter as $\Lambda=\Lambda_0+\epsilon_* \, \Lambda_1+\cdots$ and $\lambda=\epsilon_* \, \lambda_1+\cdots$ where in the last expansion we have started at  order  ${\cal O}(\epsilon_*)$ since in the absence of D7-branes there are no worldvolume fields. 
In an $\epsilon_*$ expansion the different fluctuations equations of motion can be written in the hierarchic way
\begin{align}
\label{eq.flucbg}EOM[\Lambda_0] & = 0 \ , \\
\label{eq.flucwv}EOM[\lambda_1] & = \epsilon_* \, \sigma[\Lambda_0] \ , \\
\label{eq.flucbr}EOM[\Lambda_1] & = \sigma[\lambda_1] \ , \\
\vdots\nonumber
\end{align}
where $EOM[\cdot]$ represents linear, possibly coupled, second order differential operators and $\sigma[\cdot]$ some linear, at most first order, differential operators acting as a source.\footnote{Notice that $EOM[\cdot]$ and $\sigma[\cdot]$ are not necessarily the same in every equation; we have omitted extra labels for the sake of clarity.}

If we had considered a D7 \emph{probe} with a trivial embedding in our setup, the equations governing the dynamics of this extra D7 would be given by \eqref{eq.flucwv} with $\sigma[\Lambda_0]=0$ (because the supergravity fields remain frozen in the probe approximation).
In the present setup supergravity and worldvolume fields fluctuate with amplitudes of the same order and their equations are naturally coupled.

We will consider only leading order effects, since the higher order in $\epsilon_*$ terms will contribute just with small corrections that will not change our conclusions.
Working at leading order means that equations \eqref{eq.flucbr}, etc. are not considered. This implies particularly that the backreaction corrections described in section \ref{sec.chargedblackbrane} do not appear explicitly in the equations we solve.

%%%%%%%%
\subsection{Restricting the fluctuating modes}

Due to the presence of non-trivial components in the background RR potentials $C^{(2)}_1$ and $C^{(2)}_2$, a  complete analysis of all the perturbations becomes very involved.
However, we can consider a consistent truncation of such perturbation modes, which includes in particular  the field we are interested in, $\cA_0$. 
We will consider that any field $\Psi$ in the background solution\footnote{In particular $\Psi$ can have tensorial indices which we are not writing, and can have the background value $\Psi=0$.} is perturbed by
\be
\Psi \to \Psi + e^{i k.x}\delta \Psi \ ,
\ee
where we can use the residual $SO(3)$ little group invariance of the solution we are perturbing to set $k\cdot x=-\omega t+ q x^3$. 
Since we will consider fluctuations in a black hole background, the related frequencies associated to the $\delta \Psi$ modes will generically be complex.
We will also have to pick up appropriate boundary conditions for the fluctuating fields at the horizon: incoming-wave boundary conditions at the horizon are known to be relevant for computing retarded correlators in the dual theory .

We will restrict our analysis to the $q=0$ case in which the little group remains unbroken. 
As a result, the fluctuating modes classify  into tensorial, vectorial and scalar $SO(3)$ modes.
As we explained in the introduction, we are mostly interested in analyzing the behavior of the fluctuation of the $\cA_0$ mode, which is a scalar under $SO(3)$. Therefore, we will focus just on the scalar fluctuations of the fields.
These include (see also Table \ref{tab.fields}) perturbations of all the scalars present in the system, and components of the vector and tensor fields spanning the $t-r$ subspace. Components of fluctuations of the metric in the $t-r$ directions, as well as its trace, must be considered as well. Counting the number of modes we have
\begin{equation*}
1\times 8\text{ (scalars)} + 2\times 5\text{ (vectors)}+1 \times 2\text{ ($2$-forms)}+4\text{ (metric)} = 24\text{ modes.}
\end{equation*}
This is still a large number of modes to analyze. 
We can make a further reduction on the number of fields by imposing that the equations of motion that do not vanish identically in the background (the ones for the three scalars $f$, $w$ and $\Phi$, Einstein equations and the equations for $H^{(3)}_3$ and $F^{(3)}_2$) do not receive corrections at first order in fluctuations. 
This implies that we will not consider fluctuations of non-vanishing background scalar fields, nor the metric. 

Furthermore, consistency requires that fluctuations of the two-form mode $B^{(2)}_2$ to cancel. 
Doing the counting we observe that now we have four scalars ($\delta B^{(2)}_0$, $\delta C^{(0)}_0$, $\delta C^{(4)}_0$ and $\delta\cA_0$) and the $t$ and $r$ components of three vectors ($\delta B^{(2)}_1$, $\delta C^{(4)}_1$ and $\delta\AR$), adding up to ten different modes which form a closed system (in principle we could have considered the $tr$ component of the two-form $\delta C^{(2)}_2$ as well, but it is pure gauge in this setup).
In appendix \ref{app.perturbations} we list the equations of motion describing these perturbations at leading order, along with technical comments.
From now on we skip the $\delta$ in the names of perturbation fields.

A schematic description of the operators dual to the fluctuating fields we consider is:
\begin{itemize}
\item[-] the axion corresponds to the $\Delta=4$ operator $F\wedge F$  in the field theory, and will not play an essential r\^ole in our discussion below;
\item[-] the NSNS vector field $B^{(2)}_1$ corresponds holographically to an operator of  dimensionality $\Delta=2+\sqrt{9+4Q_f}=5+{\cal O}(Q_f)$ and sits in the supermultiplet with the term $\mathrm{Tr}(\bar {\cal W}_{\dot\alpha} {\cal W}_{\beta}{\cal W}^{\beta})+ \cdots$;
\item[-] the fields $C^{(4)}_1$ and $\AR$ correspond to a transformed basis for two vector fields holographically dual to $\Delta_{J_R}=3$ and $\Delta=7$ operators. The operator $J_R$ generates the $U(1)_R$ symmetry \cite{Cassani:2010uw,Gauntlett:2010vu}; 
\item[-] the operator dual to the scalar field ${\cal A}_0$ is sensitive to the value of $\sigma$. For $\sigma=1$ (anti-D7-brane case) the scalar is dual to an operator of dimension $\Delta=6+Q_f$. For  $\sigma=-1$ (D7-brane case) the scalar is dual to an operator of dimension $\Delta=2+Q_f$ \cite{Cotrone:2012um}. In the probe approximation its mass sits on the $AdS_5$ BF bound and the scalar is dual to \cite{Kruczenski:2003be}
\be
{\cal O} =  q^{\dagger\alpha} \sigma_{\alpha \beta} q^{\beta} \ ,
\ee
where $q$ is the doublet of $SU(2)_R$ squarks and $\sigma_{\alpha\beta}$ are Pauli matrices. The scalar operator is a vector of $SU(2)_R$, which is dubbed ``R-spin'' in \cite{Ammon:2011hz}.
The case  $\sigma=-1$ is the one we will focus on.
\end{itemize}

%%%%%

In appendix \ref{app.perturbations} we have redefined the gravity fields dual to the $\Delta=5,7$ operators into gauge-invariant combinations (which we dubbed $\eta_B$ and $\eta_C$ respectively, but see the comments around \eqref{eq.realCop} and \eqref{eq.realBop}), which allowed us to decouple completely the $\epsilon_*^0$ equations.
The scalar dual to the $\Delta=2$ operator  (dubbed $\eta_{\cal A}$) is  sourced by the former gauge invariant combinations (as in equation \eqref{eq.flucwv}). 

Mixing of perturbations' equations of motion is a generic situation in holographic systems.
In \cite{Kaminski:2009dh} this situation was studied in generality and it was found that the generalization of the two-point function's prescription of \cite{Policastro:2002se} consisted, schematically, on the expression
\be\label{eq.greens}
G_R \propto B \cdot A^{-1} \ ,
\ee
where the subindex ${}_R$ indicates that we are talking about the retarded two-point function (which is imposed by fixing incoming wave boundary conditions at the horizon \cite{Policastro:2002se}), $B$ is schematically a matrix of normalizable modes (dual to vevs of the operators \cite{Kaminski:2009dh}) and $A$ a diagonal matrix of non-normalizable modes (dual to sources of the operators). Focusing on the fields  $ C_0^{(0)}\equiv  \chi$, $ \eta_{\cal A}$, $ \eta_B$ and $\eta_C$ we have in our case
\be
G_R = \begin{pmatrix}
V_{\chi\leftarrow\varphi_\chi} & V_{\chi\leftarrow\varphi_{\cal A}} & V_{\chi\leftarrow\varphi_B} & V_{\chi\leftarrow\varphi_C} \\
V_{{\cal A}\leftarrow\varphi_\chi} & V_{{\cal A}\leftarrow\varphi_{\cal A}} & V_{{\cal A}\leftarrow\varphi_B} & V_{{\cal A}\leftarrow\varphi_C} \\
V_{B\leftarrow\varphi_\chi} & V_{B\leftarrow\varphi_{\cal A}} & V_{B\leftarrow\varphi_B} & V_{B\leftarrow\varphi_C} \\
V_{C\leftarrow\varphi_\chi} & V_{C\leftarrow\varphi_{\cal A}} & V_{C\leftarrow\varphi_B} & V_{C\leftarrow\varphi_C} 
\end{pmatrix} \cdot \begin{pmatrix}
\varphi_\chi^{-1} & 0 & 0 & 0 \\
0 & \varphi_{\cal A}^{-1} & 0 & 0 \\
0 & 0 &\varphi_B^{-1} & 0 \\
0 & 0 & 0 & \varphi_C^{-1}  
\end{pmatrix} \ ,
\ee
where $\varphi_X$ is the source of the operator dual to the field $X$, and $V_{X\leftarrow \varphi_Y}$ is proportional to the vev of the operator dual to the field $X$ when \emph{only} $\varphi_Y$ is  non-zero. This, together with the ingoing-wave condition at the horizon, determines the boundary conditions used to calculate the $V_{X\leftarrow\varphi_Y}$, which are presented in the next section.% and defined in \eqref{eq.greenscomponents}.

 The position of the quasinormal modes is a property of the retarded two-point function as a matrix, as the poles in a meromorphic expansion, and therefore all the $V_{X\leftarrow \varphi_Y}$ terms, are described by the same set of QNMs.

In the former discussion we have not considered possible divergences arising in \eqref{eq.greens}, i.e., we have assumed that a holographic renormalization procedure has been made such that we can identify the matrix $B$ in that expression with the matrix of normalizable modes. 
We have not proved that a proper cancellation occurs when explicit counterterms are added, especially with the lack of an asymptotically locally $AdS$ spacetime. Here we  assume this is the case. 
Indeed, background subtraction is enough to obtain the correct thermodynamical relations in which the free energy is obtained via the on-shell action, which is UV divergent. 
That the normalizable mode is the relevant factor to calculate the two-point function is a well established fact in holography \cite{Skenderis:2002wp}, even in the presence of irrelevant operators \cite{vanRees:2011fr}.\footnote{Due to the structure of the effective perturbation action, counterterms contribute only to the hermitian part of the $G_R$ matrix. Therefore, for the two-point function the anti-hermitian combination
\be
\rho(\omega) = i \left( G_R(\omega) - G_R(\omega)^\dagger \right) \ ,
\ee
should not be sensitive to the addition of such counterterms. In \cite{Kaminski:2009dh}  it was shown that the spectral function matrix can be defined at any radius, contrary to the two-point function which is defined at the boundary. Since by construction all fields are regular at the horizon, the definition of the spectral function at the horizon proves  its finiteness. The regularized hermitian part can be obtained by use of Kramers-Kr\"onig relations.}

%%%%%

\subsection{The retarded Green's function: numerical results}

The fluctuation of the axion at leading order \eqref{eq.axionfluc} was already studied in the literature \cite{Starinets:2002br,Musiri:2003rs}, and the quasinormal modes were found to be at $\omega=2\pi T n(\pm1-i)$ with $n>0$ an integer. 
In particular the frequency-function $\frac{V_{\chi\leftarrow \varphi_\chi}}{\varphi_\chi}$ can be found in \cite{Starinets:2002br}.
As we will see below, the rest of the components $\frac{V_{X\leftarrow \varphi_Y}}{\varphi_Y}$ we analyze do not show QNMs at $\omega=2\pi T n(\pm1-i)$, even when the QNMs are shared by all components of the Green's function. The reason for this is the $\epsilon_*$ expansion. If we had included higher order corrections, the couplings between the equations of motion would have given rise to the presence of QNMs at precisely these values (plus an $\epsilon_*$ correction). In different words, the residue of these poles for the other $\frac{V_{X\leftarrow \varphi_Y}}{\varphi_Y}$ components is subleading in $\epsilon_*$.

\begin{figure}[tb]
\begin{center}
\includegraphics[scale=0.6]{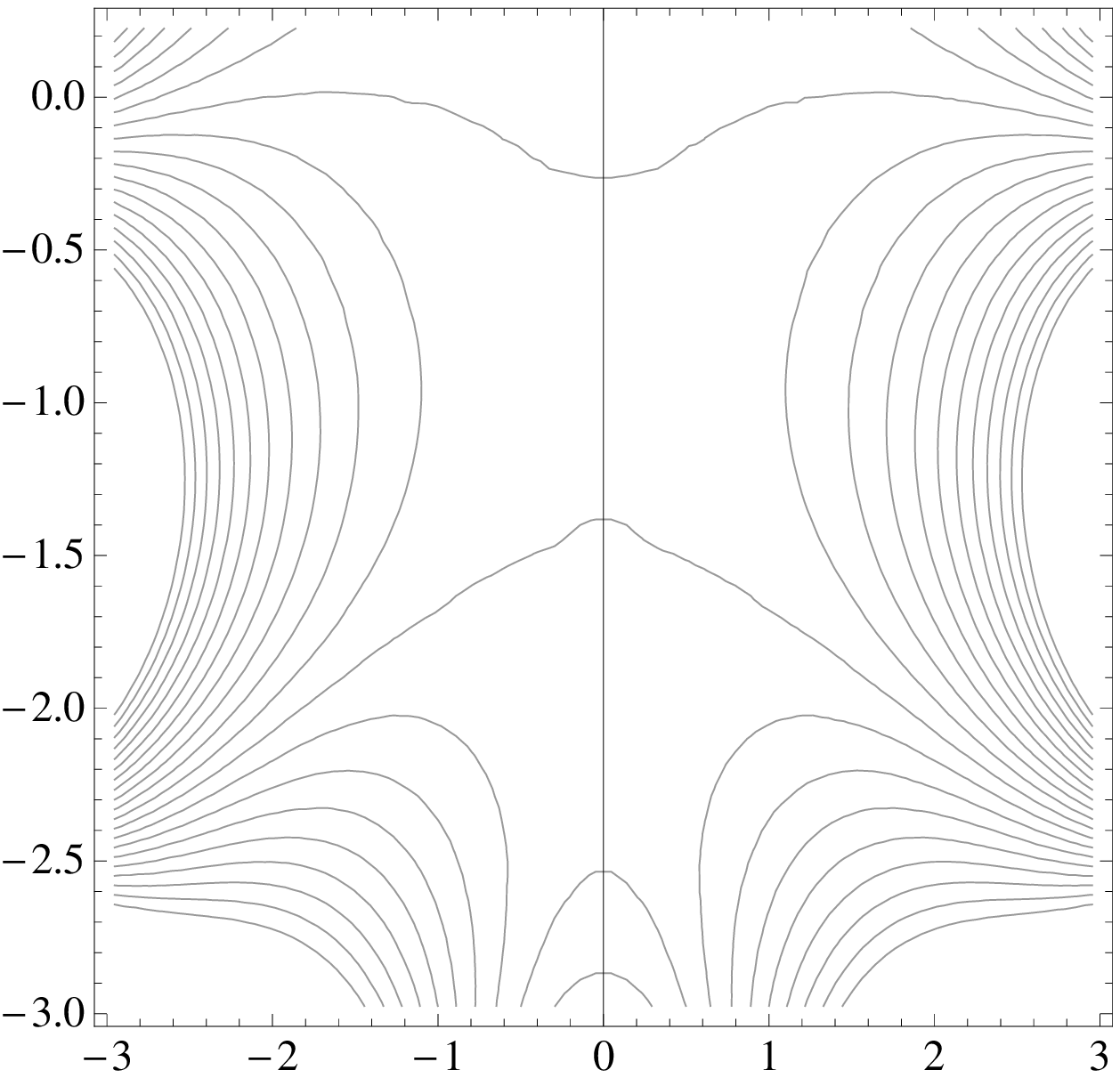}
\includegraphics[scale=0.6]{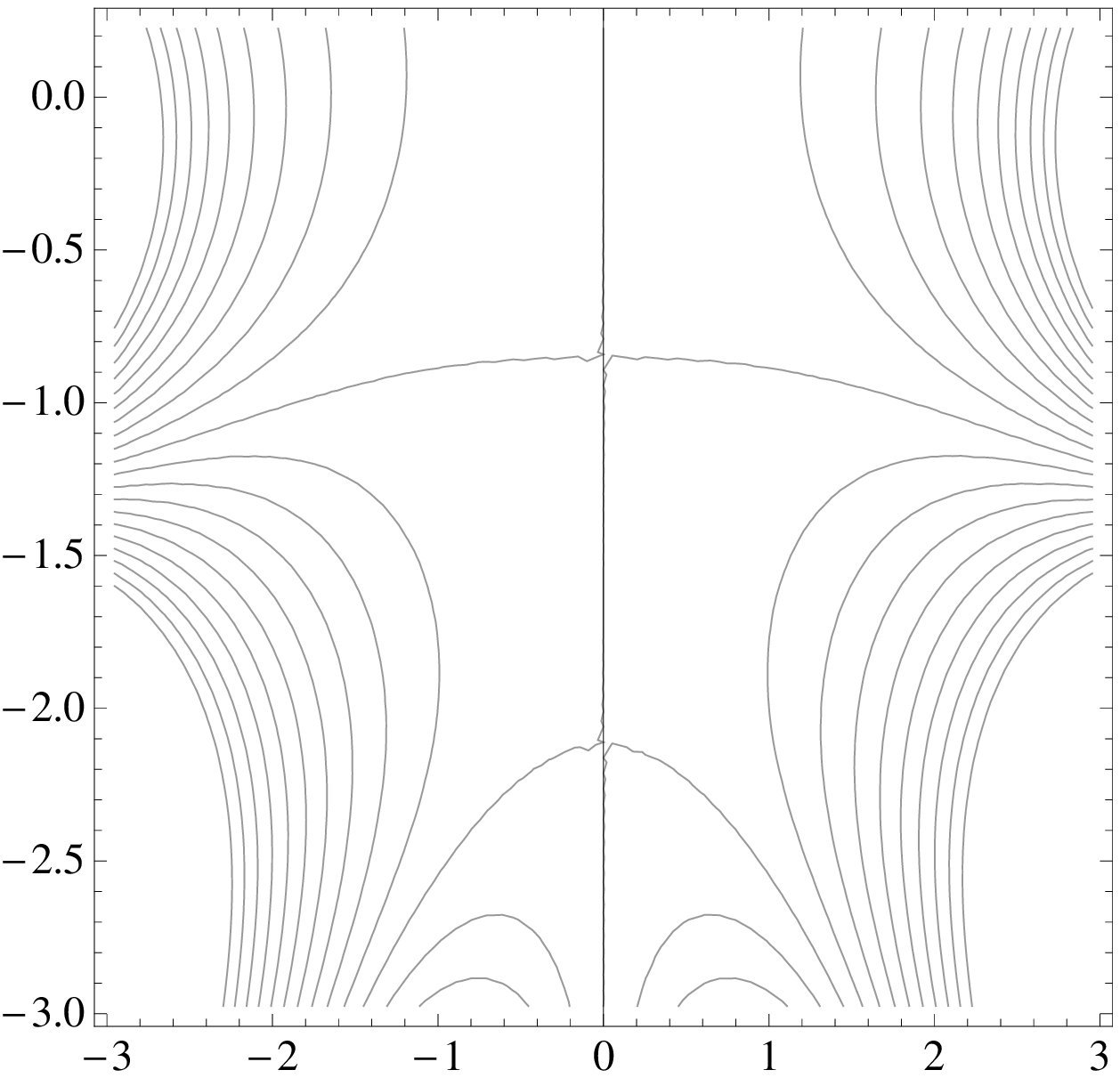}
\caption{Contour plots of the real (left) and imaginary (right) parts of $\frac{V_{C\leftarrow \varphi_C}}{\varphi_C}$  in the complex $\omega/r_h$ frequency plane. The real (imaginary) part is even (odd) under $\omega \to -\bar\omega$.}\label{fig.etaC}
\end{center}
\end{figure}

The only physical parameter appearing in the leading order equation of motion for the fluctuation $\eta_C$, \eqref{eq.etaC}, is the temperature, which is encoded in the radius of the horizon $r_h = \pi T + {\cal O}(\epsilon_*)$. 
This means that at leading order in $\epsilon_*$ we can work with the dimensionless ratio $\omega/r_h$.
Taking $r_h=1$ for convenience we present in figure \ref{fig.etaC} a plot of the real and imaginary parts of $\frac{V_{C\leftarrow \varphi_C}}{\varphi_C}$ in the negative imaginary frequency plane (no non-analyticities were observed in the upper half plane), following appendix \ref{app.perturbations}.
As explained in the appendix, the contour plot shown in figure \ref{fig.etaC} is valid up to some contact terms which, in particular, will not introduce any non-analyticities. Therefore we can say that at leading order in the backreaction parameter, for frequencies not much larger in  norm than  $\pi T$, there are no QNMs in the propagator of $\eta_C$.

\begin{figure}[tb]
\begin{center}
\includegraphics[scale=0.6]{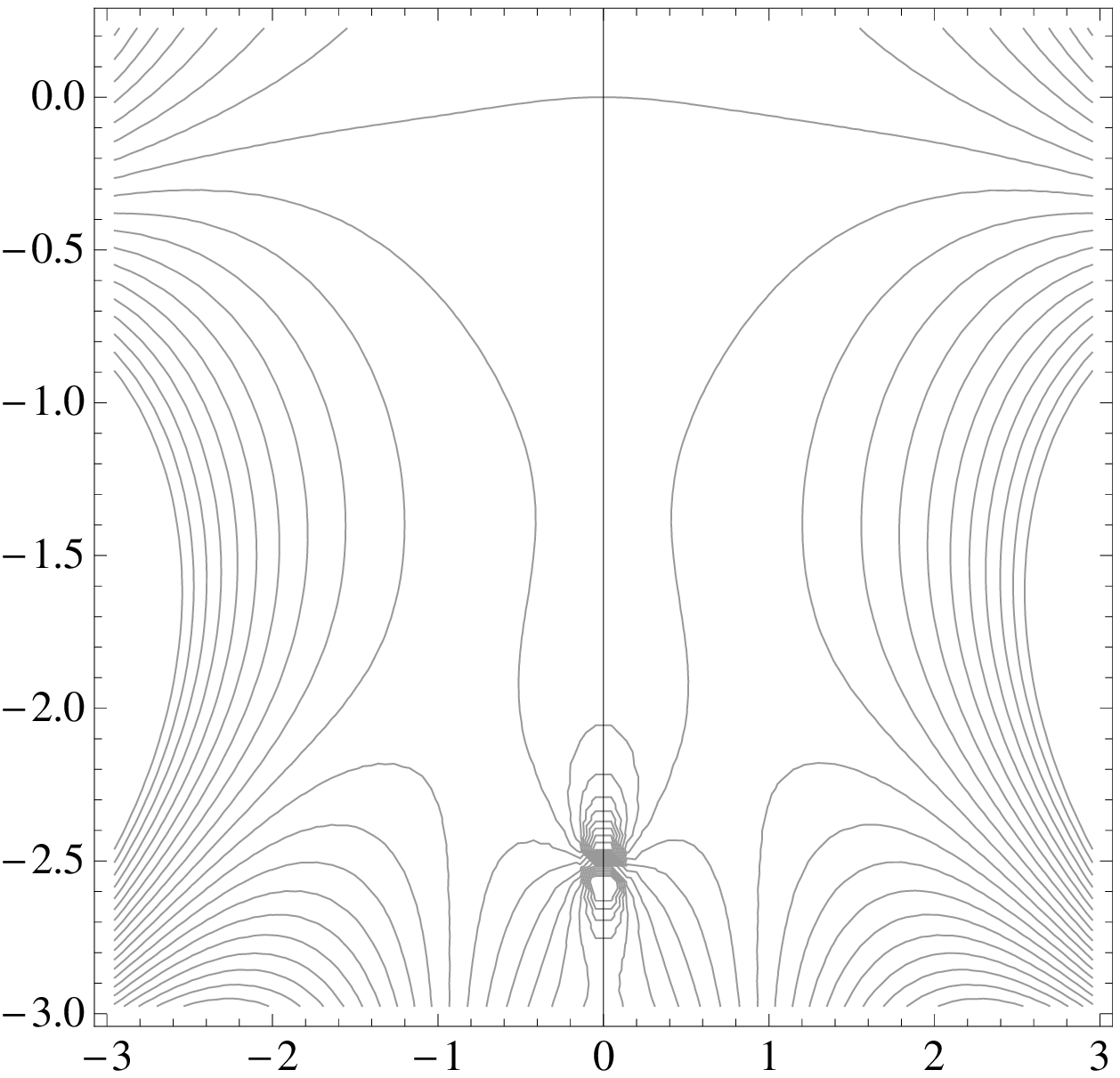}
\includegraphics[scale=0.6]{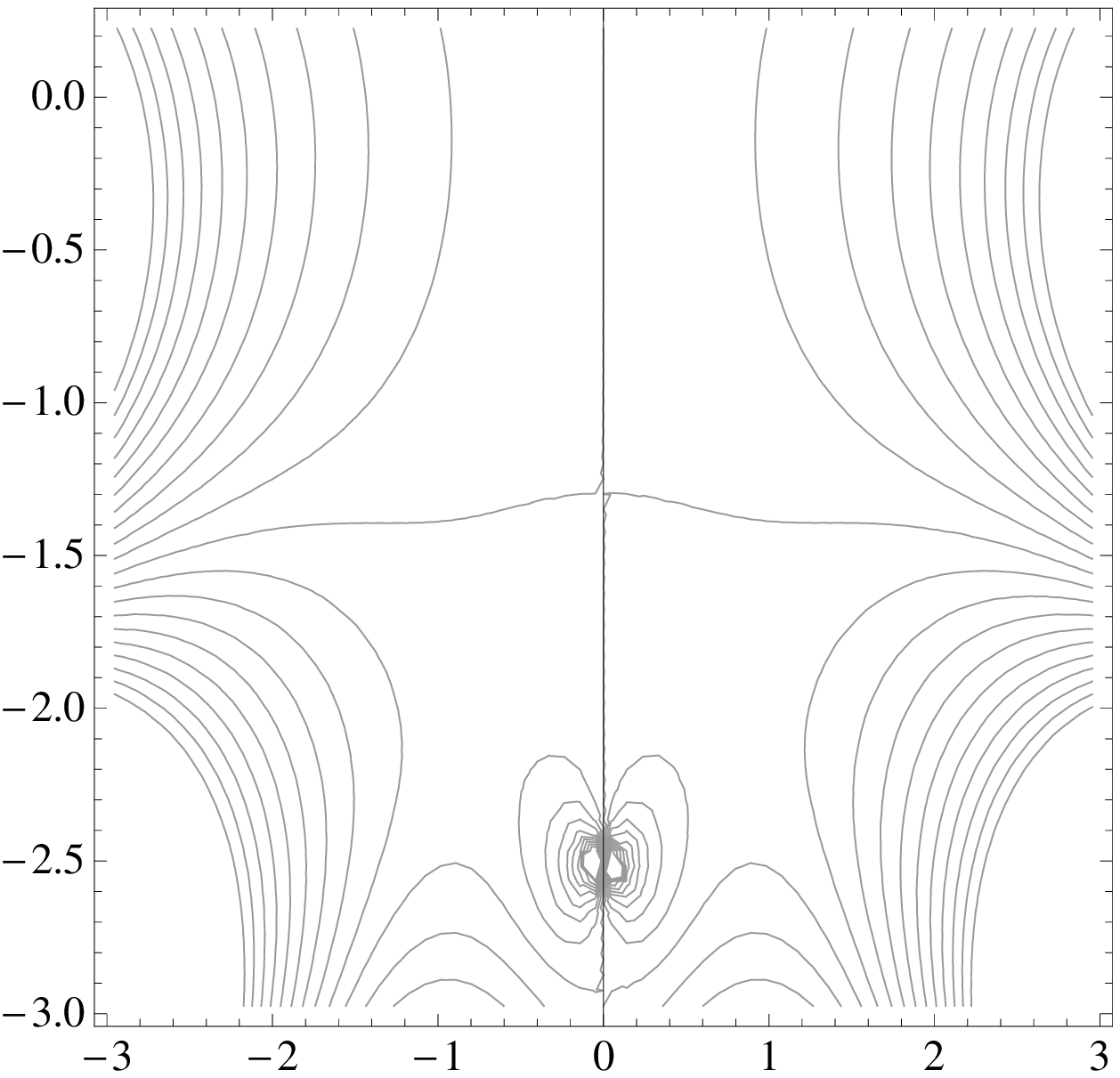}
\caption{Contour plots of the real (left) and imaginary (right) parts of $\frac{V_{B\leftarrow \varphi_B}}{\varphi_B}$  in the complex $\omega/r_h$ frequency plane. The real (imaginary) part is  even (odd) under $\omega \to -\bar\omega$.}\label{fig.etaB}
\end{center}
\end{figure}

Similar comments hold for $\frac{V_{B\leftarrow \varphi_B}}{\varphi_B}$, which we present in figure \ref{fig.etaB}, up to contact terms, for the negative imaginary frequency plane (no non-analyticities were observed in the upper half plane). 
We see now the existence of a quasinormal mode in the imaginary axis for $\omega/r_h \approx -2.5 i$. This mode is purely damped and gapped, therefore it does not correspond to a hydrodynamic mode. The fact that this QNM is not observed in the perturbation for the axion is, once again, an effect of the $\epsilon_*$ expansion. On the other hand, we will see below how this mode appears in other components of the Green's function.

Now we come to $\frac{V_{{\cal A}\leftarrow \varphi_{\cal A}}}{\varphi_{\cal A}}$. As seen in \eqref{eq.etaA} there are two dimensionful physical parameters: the temperature (related to $r_h$) and the charge density (related to $r_d$).
 As we argued in the previous section, our solution is valid only for temperatures larger than the charge density per quark, so it is natural to scale dimensionful quantities with the temperature, namely $\omega/r_h$ and $r_d/r_h\sim n_q^{1/3}/T$.

\begin{figure}[tb]
\begin{center}
\includegraphics[scale=0.6]{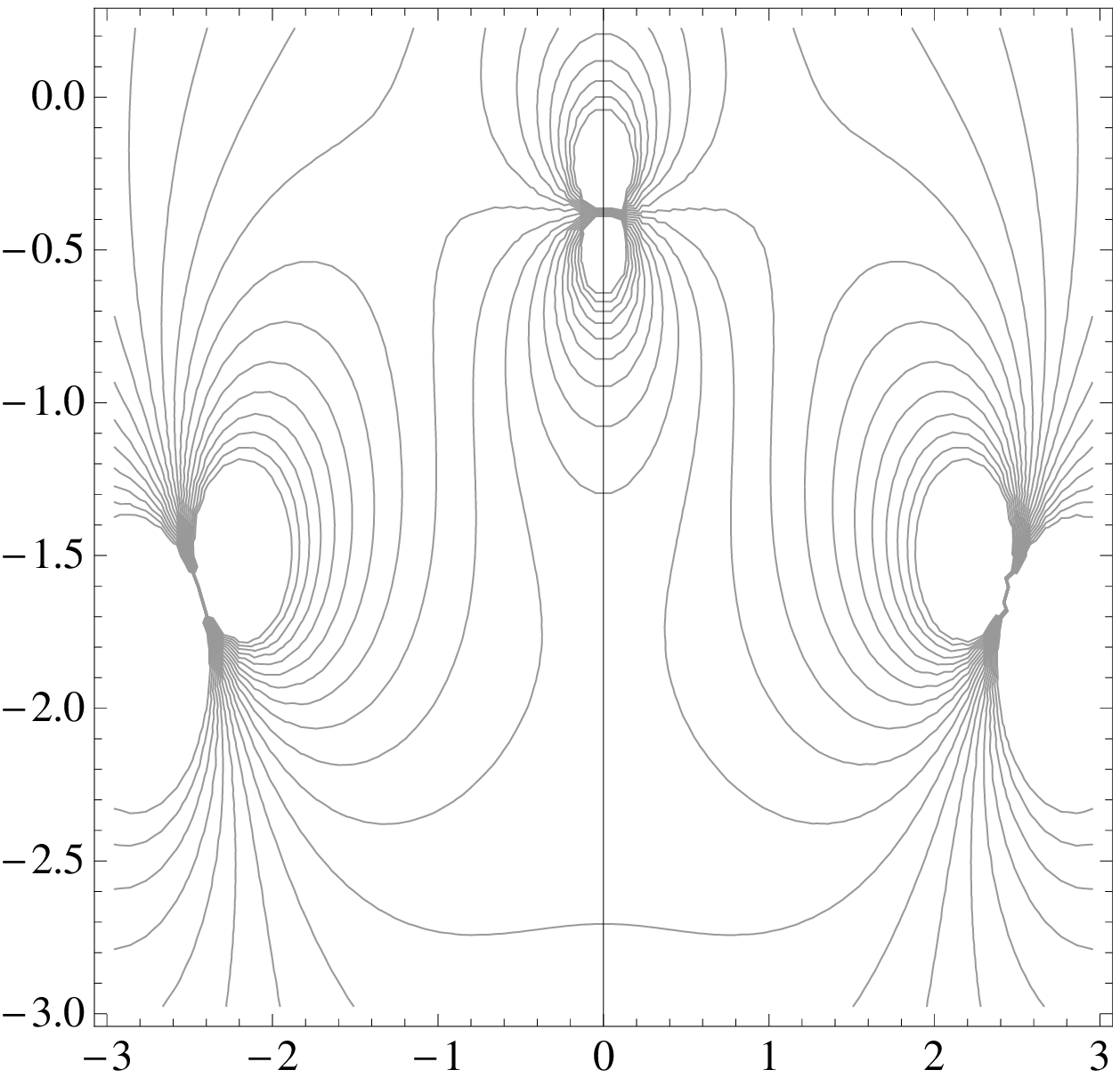}
\includegraphics[scale=0.6]{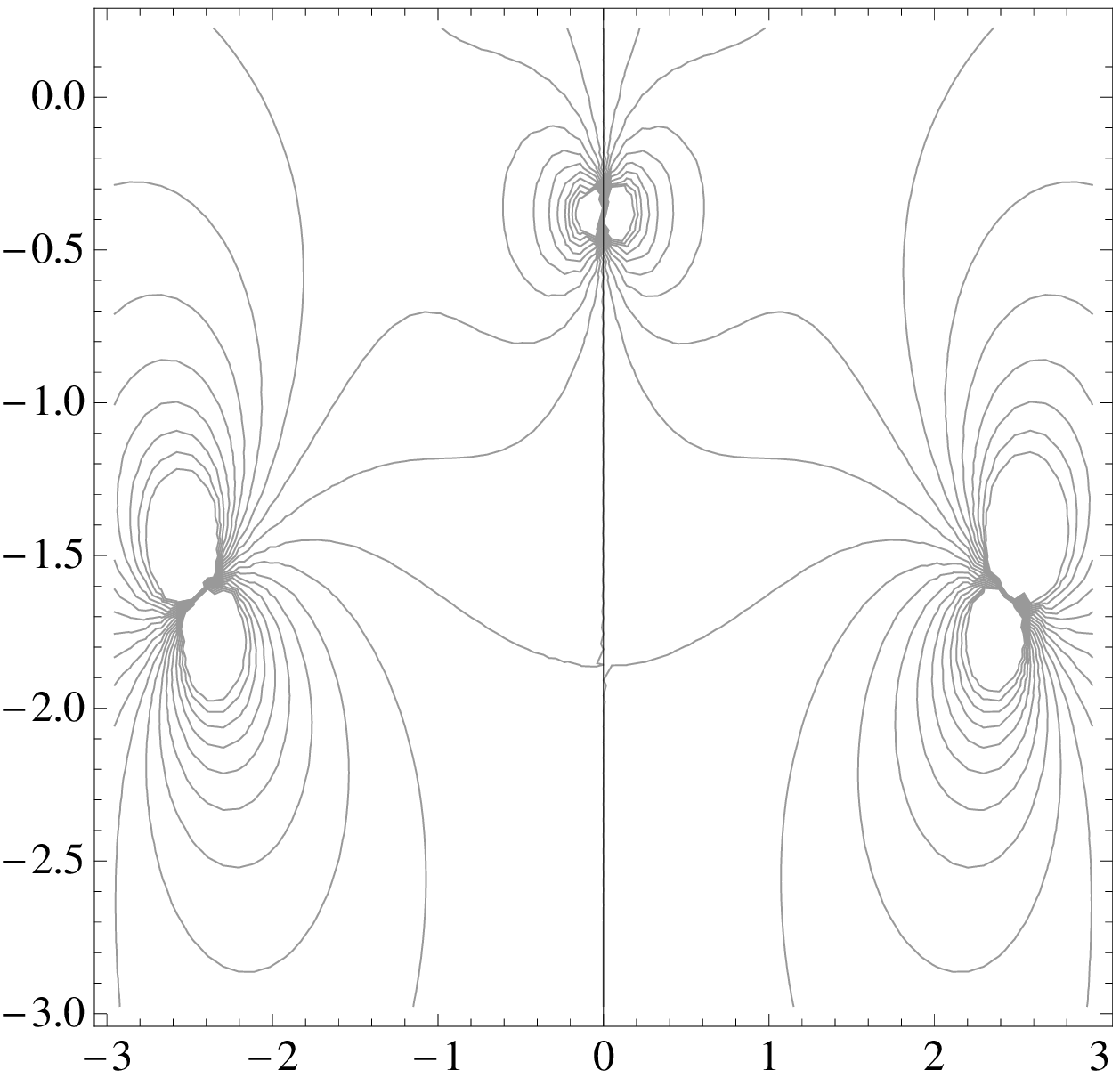}
\caption{Contour plots of the real (left) and imaginary (right) parts of $\frac{V_{{\cal A}\leftarrow \varphi_{\cal A}}}{\varphi_{\cal A}}$ in the complex $\omega/r_h$ frequency plane for $r_d/r_h=1.4$. The real (imaginary) part is  even (odd) under $\omega \to -\bar\omega$.}\label{fig.etaA}
\end{center}
\end{figure}

\begin{figure}[tb]
\begin{center}
    \includegraphics[scale=0.5]{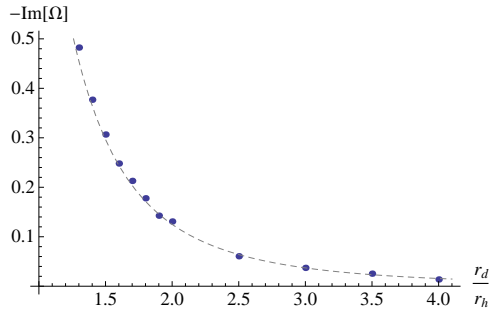}
  \caption{Position of the purely damped mode in the $\omega/r_h$ imaginary axis. The dashed line corresponds to $(r_d/r_h)^{-3}$. \label{fig.QNM0}}
  \end{center}
\end{figure}

In figure \ref{fig.etaA} we show a typical contour plot for $\frac{V_{{\cal A}\leftarrow \varphi_{\cal A}}}{\varphi_{\cal A}}$ where three QNMs can be observed. 
One of the QNMs corresponds to a purely damped mode and presents a gap (its imaginary part is non-zero).
In figure \ref{fig.QNM0} we show how the purely damped mode behaves as we change the charge density relative to the temperature. 
In this plot we show the curve $(r_d/r_h)^{-3}$ as a guide to the eye, however we find a remarkable agreement between this line and the actual position of the damped mode, although we did not find an analytic explanation for it. 
Assuming this mode is actually located at
\be\label{eq.interestingQNM}
\Omega=- \frac{r_h^3}{r_d^{3}} \, i \ ,
\ee
would imply that the mode will never become unstable by crossing to the upper-half plane, it will just approach the real axis as the radius of the horizon is small respect to the charge density. This is  what was found in \cite{Ammon:2011hz}, where a zero sound mode is found at the origin for the $r_h=0$ case ($T=0$ in the probe limit).

\begin{figure}[tb]
\begin{center}
\includegraphics[scale=0.5]{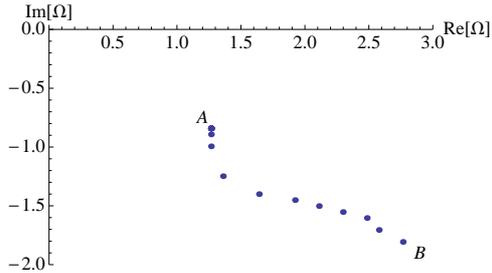}
\caption{Position in the $\omega/r_h$ complex plane of the first quasinormal mode with non-null real part (by symmetry there is another QNM related to the one showed in the graph with opposite real part) from $r_d/r_h=0.6$ (point A in the graph) to $r_d/r_h=1.6$ (point B). At lower values of $r_d/r_h<0.6$ the position of the quasinormal mode does not change perceptively.}\label{fig.QNM1}
\end{center}
\end{figure}

The behavior of the oscillatory modes (those with a real part in the complex frequency plane) with the charge density is given in figure \ref{fig.QNM1}, where it is observed that increasing the ratio $r_d/r_h$ pushes the modes deeper into the complex frequency plane (where the frequency is given in units of the radius of the horizon).

\begin{figure}[tb]
\begin{center}
\includegraphics[scale=0.6]{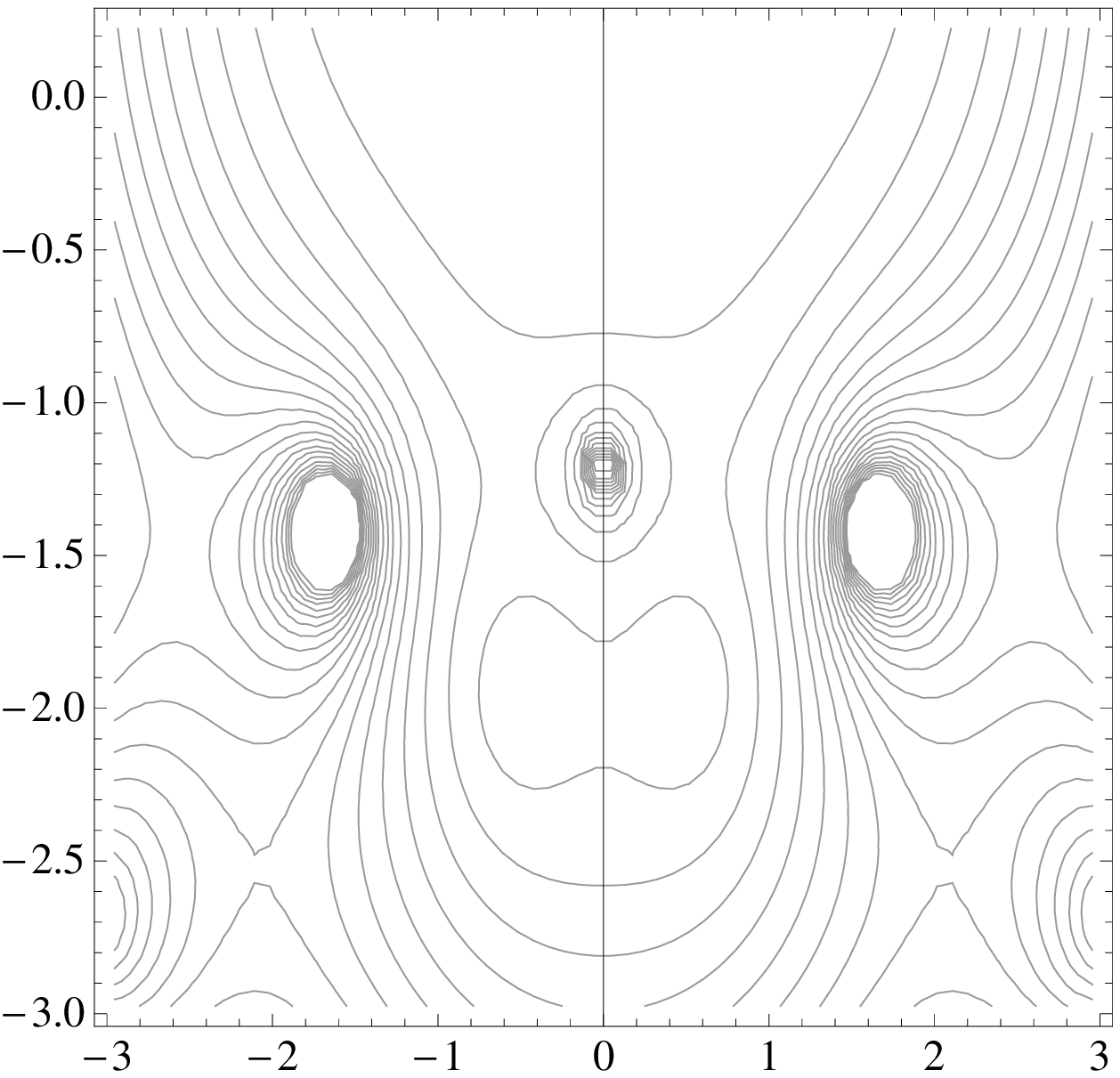}
\includegraphics[scale=0.6]{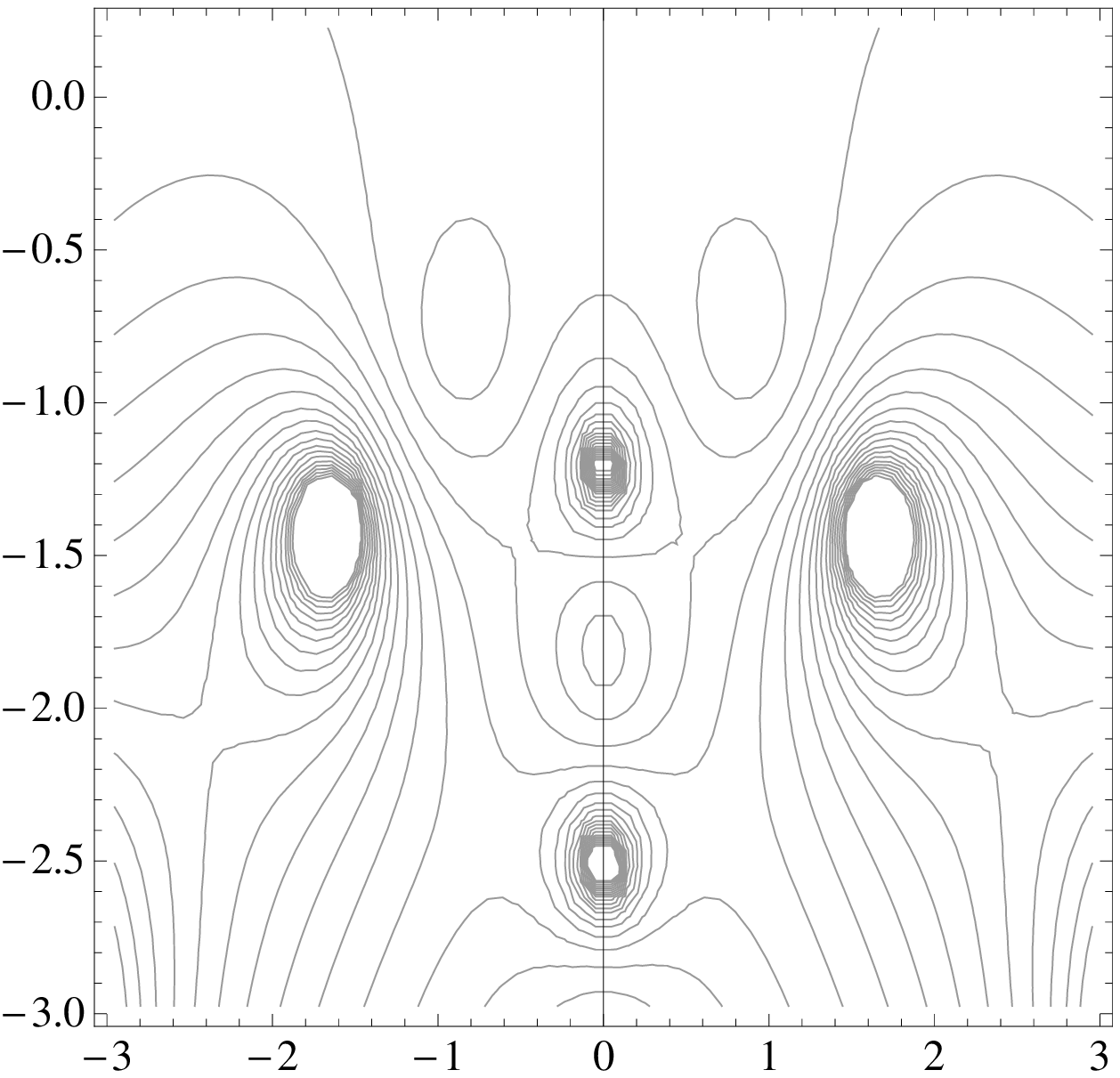}
\caption{Absolute value of $\frac{V_{{\cal A}\leftarrow \varphi_X}}{\varphi_X}$. \emph{Left:} case (a) discussed in the text, with $X=C$. \emph{Right:} case (b) discussed in the text, with $X=B$. In these figures $r_d/r_h=1$. \label{fig.etaACetaAB}}
\end{center}
\end{figure}

Sourcing explicitly the operators dual to $\eta_B$ and $\eta_C$  sources  a vev for the operator dual to $\eta_{\cal A}$, as can be seen from equation \eqref{eq.etaA},  where the presence of non-trivial $\eta_B$, $\eta_C$ would trigger a non-trivial profile for ${\cal A}$, even if we keep the boundary condition $\varphi_{\cal A}=0$. We indeed observe this when we integrate the equations of motion with boundary conditions (a) $\varphi_{\cal A}=\varphi_B=0$ and $\varphi_C=1$ or (b) $\varphi_{\cal A}=\varphi_C=0$ and $\varphi_B=1$. 
In the case (a) the contour plot for  $\frac{V_{C\leftarrow \varphi_C}}{\varphi_C}$ is given by figure \ref{fig.etaC}, and the one for  $\frac{V_{{\cal A}\leftarrow \varphi_C}}{\varphi_C}$ in figure   \ref{fig.etaACetaAB} (left). We can see quasinormal modes whose positions depend on $r_d/r_h$. The position of these quasinormal modes is the same as in figure \ref{fig.etaA}.
In the case (b) again $\frac{V_{B\leftarrow \varphi_B}}{\varphi_B}$ is given by figure \ref{fig.etaB}, and the contour plot for $\frac{V_{{\cal A}\leftarrow \varphi_B}}{\varphi_B}$ (see figure \ref{fig.etaACetaAB} (right)) presents quasinormal modes with positions coincident with the quasinormal modes observed both in figure \ref{fig.etaA} and \ref{fig.etaC}.
These coincident QNMs are not surprising, as we commented already, since these equations are coupled  and QNMs are shared by the different components of the matrix of Green's functions. This is the dual to the mixing of operators in the field theory side, such that sourcing one single operator inevitably implies a vev for more than one operator. 

Summarizing the results presented in this section, we observe that all the QNMs at leading order in $\epsilon_*$ are in the lower complex plane, and therefore are stable modes. Going to next order in $\epsilon_*$ would require to solve equations of motions where the corrections to the background solution found in section \ref{sec.chargedblackbrane} appear explicitly, however such a correction could modify the position of the quasinormal modes only perturbatively in $\epsilon_*$, and would not trigger an instability. 
The only case where such a correction might have an important effect would be in the $r_h/r_d \to0$ case, where \eqref{eq.interestingQNM} points to a mode sitting at the origin of frequencies. However, this situation is out of the regime of validity of our solution.

%%%%%%%%
%%%%%%%%%

%%%%%%%%%%%%%%%%%%%
%%%%%%%%%%%%%%%%%%%%%%

\section{Conclusions}\label{sec:conclusions}
We have expanded the study of the charged D3-D7 black hole started in \cite{Bigazzi:2011it} providing a new solution for generic value of the parameter $r_d$. 
The solution is dual to ${\cal N}=4$ SYM (and quiver generalizations) with massless dynamical flavors at finite temperature and baryon density.
Let us discuss in turn two features of this solution.

\subsection{Breaking of the perturbative solution and of the probe approximation}

From equation \eqref{dnq} (using $n_q=N_c n_B$ and specializing to the flavored ${\cal N}=4$ SYM case), we observe that the dimensionless quotient $r_d/r_h$ is a measure of the charge (baryon) density per flavor, $n_{B/q}=\frac{n_B}{N_f}$, over temperature
\be
\frac{r_d^3}{r_h^3} = \frac{8\, n_{B/q}}{\sqrt{\lambda_*}T^3} + {\cal O}(\epsilon_*) \ .
\ee
From this formula it follows that, since $r_d/r_h$ can be of order one in our solution, the latter can describe effects due to large charge density per flavor (order $\sqrt{\lambda_*}T^3$).

Nevertheless, the overall value of the charge density is still ``small'', in the sense that it is not sufficient to modify at order one the IR physical observables.
In fact, the latter are of the generic form $O \sim O_{{\cal N}=4}\left[ 1+\epsilon f(r_d/r_h) + ... \right]$.
As we have seen in section \ref{sec.extremal}, e.g. for the temperature, the flavor corrections $\epsilon f(r_d/r_h)$ are of order one when $(r_d/r_h)^3 \sim 1/\epsilon$, but, as expected, this value makes higher order corrections in $\epsilon$ relevant, thus spoiling the reliability of the solution.

From the gravitational side this is expected since the gravitational force would not be able to compensate the electric repulsion in a small black hole. 
It is not clear to us what to expect in the field theory side. A transition to a different phase, like the spatially modulated phases encountered for the dual of the $AdS$-RN black hole \cite{harvey} would appear if instabilities are found at finite momentum, which we have not covered in the present work. 
%In the field theory side one could expect, by comparison with similar systems, a transition to a different phase, like the spatially modulated phases encountered for the dual of the $AdS$-RN black hole \cite{harvey}. 

Thus, while the solution can correspond to large charge with respect to the temperature, it does not describe in the IR an (almost) zero temperature system, which would require a solution at all orders in $\epsilon f(r_d/r_h)$.
The existence and the nature of such solution remains an extremely interesting open problem.

\vspace{0.5cm} 
The breakdown of the perturbative solution at zero temperature implies the breakdown of the charged probe approximation as well.
In fact, the probe approximation requires that the original background is not significantly affected (or it is un-affected for a long time) by the addition of a small perturbation due to flavor branes (i.e. $\epsilon f(r_d/r_h) \ll 1$).
But in the present case the original background does not admit a small perturbation at zero temperature: either the perturbed system is at non-zero temperature (and ``far'' from zero temperature in a qualitative way, as we have seen in this paper), or the flavor contribution must be dominant near the origin of $AdS$ in order to have a zero or almost zero temperature (the corresponding solution being still unknown).

This breakdown of the charged probe approximation is actually already known. For example, in \cite{Hartnoll:2009ns} it is pointed out that for the probe approximation to make sense, the stress-energy tensor of the probe must be subleading with respect to the one of the background, otherwise one has to take into account the brane backreaction.
In a zero temperature background and for a charged probe, this condition cannot be satisfied all the way up to the deep IR region,\footnote{Condition (3.27) in \cite{Hartnoll:2009ns} is equivalent to the condition for the validity of our solution $(r_d/r_h)^3 \ll 1/\epsilon$.} unless there is some other mass scale, such as the temperature or the flavor mass. 
In the latter cases, the probe approximation is valid as long as the charge density is not parametrically larger (in $N_c$) than these extra scales.

\subsection{Stability}
 
With the aim of finding a physical instability we have checked the matrix of susceptibilities of the system, which turned out to be parametrically stable, and quasinormal modes of a set of perturbations of the system, showing no indication of a physical instability either. We comment further on the quasinormal modes below.

\paragraph{Comparison of the fluctuation calculation with the probe limit.} In equations \eqref{eq.flucbg}--\eqref{eq.flucbr} we have sketched how the equations for the fluctuations organize  when a perturbative expansion in $\epsilon_*$ is implemented.
In the probe approximation fluctuations of the supergravity fields are not allowed and $\Lambda_0=0$.
This  satisfies identically equation \eqref{eq.flucbg} and transforms equation \eqref{eq.flucwv} into the probe fluctuation equation $EOM[\lambda_1]=0$.
A non trivial solution to these equations will source fluctuations of the supergravity fields $\Lambda_1$ via equation \eqref{eq.flucbr}, which are first order in $\epsilon_*$, which is a depreciable correction to the background value in the probe limit.
It is only after a large time of order $t\sim t_0/\epsilon_*$ ($t_0$ being order one) \cite{Karch:2008uy} that the correction $\Lambda_1$ would not be small compared to the background value and the backreaction of the probe onto the geometry needs to be considered.

 So we see how the $\epsilon_*$ expansion of the fluctuations organizes itself to take into account the backreaction of the geometry, and how can one link the results to the probe approximation case.
 The probe calculation ceases to be valid at $t\sim t_0 / \epsilon_*$ where perturbations of the probe on the horizon must be considered: energy is injected into the system and eventually it will be absorbed by the black hole, which will increase in size. 
% This effect is reflected in our solution in the $\epsilon_*$ corrections to the metric functions and can be understood physically with the  computations for the temperature and entropy of the system (see the thermodynamics section of \cite{Bigazzi:2009bk} for a specific discussion).

In the specific case of the fluctuation of the field ${\cal A}_0$ (see section \ref{subsec.A}), the fact that $\eta_B$ and $\eta_C$ source the equation of motion for $\eta_{\cal A}$ points to the fact that the source will behave differently when embedded in the original background or in the perturbed one. However, if we turn off the source of the irrelevant operators that perturb the background, the probe will be insensitive to these changes (to leading order in $\epsilon_*$), and we obtain the same equation of motion one would have obtained from the probe calculation at finite temperature.

\paragraph{Operator mixing and the two-point function.}

The equations of motion for the fluctuations we have focused on consist of a set of coupled, linear, second order differential equations. 
In section \ref{sec.perturbations} we have expanded these equations in a power series in $\epsilon_*$, since our background is only valid up to first order in that expansion. 
This allowed us to decouple the $\epsilon_*^0$ equations for the fields dual to the $\Delta=5,7$ operators, however the field dual to the $\Delta=2$ operator, itself of order $\epsilon_*$, is  sourced by the former ones. 

\vspace{0.5cm} 
Equation \eqref{eq.greens} describes the mixing of the operators in the Green's function. We have given the values of $\frac{V_{{\cal A}\leftarrow \varphi_{\cal A}}}{\varphi_{\cal A}}$, $\frac{V_{B\leftarrow \varphi_B}}{\varphi_B}$ and  $\frac{V_{C\leftarrow \varphi_C}}{\varphi_{C}}$  in figures \ref{fig.etaA}, \ref{fig.etaB} and \ref{fig.etaC} respectively, and $\frac{V_{\chi\leftarrow \varphi_\chi}}{\varphi_\chi}$ can be found in \cite{Starinets:2002br}.
The off-diagonal terms, which describe the mixing of operators, $\frac{V_{{\cal A}\leftarrow \varphi_B}}{\varphi_B}$ and  $\frac{V_{{\cal A}\leftarrow \varphi_C}}{\varphi_{C}}$ correspond to the values in figure \ref{fig.etaACetaAB}. We have observed that the position of the QNMs in the off-diagonal two-point components coincide with the positions of the QNMs in the two correspondent diagonal components, as expected from operator mixing considerations.
We have not described all of the $V_{X\leftarrow \varphi_Y}$ components, since we were mainly interested in the gapped, purely damped mode described in figure \ref{fig.QNM0}. We do not expect that the inclusion of the rest of the components will add any relevant physical information. The main interest of considering all the components would be to check the relations that are imposed on the two point matrix function by  time reversal symmetry
\be
G_R(-\omega) = G_R(\omega)^* \ , \qquad G_A(\omega)=G_R(\omega)^\dagger \ .
\ee

\paragraph{Absence of overdamped modes.}

The fluctuation of the backreacted system shows no sign of instability at zero momentum. At finite temperature the QNMs have a negative imaginary part and  $\epsilon_*$ corrections, being parametrically small, cannot push these modes to the positive half-plane. The reason for this is that our solution, as stated several times now, holds  only when the temperature dominates over the charge of the system, or when the two effects are of the same order.

We have focused in a purely damped mode with a gap at finite temperature apparently given by equation \eqref{eq.interestingQNM}. In the probe case at zero temperature this mode has been observed in \cite{Ammon:2011hz}.
For this mode to be present in the probe calculation it is crucial that fluctuations of worldvolume fields effectively see an $AdS_2$ spacetime in the IR, which is associated to extremal black holes (and therefore zero temperature solutions).
In the present work we expect the zero temperature setup to be described not by $r_h=0$, but an extremal black hole with finite horizon radius, whose near-horizon geometry would be $AdS_2\times \mathrm{R}^3$, and this would imply that the gap is non vanishing at zero temperature.

We have shown that the low temperature regime is beyond the applicability of our solution, so the former comments are just speculative.
Even if we had a $T=0$ solution at hand it is possible that no instability would appear. The reason is that in \cite{Cotrone:2012um} it was shown that backreaction corrects the mass of the field dual to the $\Delta=2$ operator to a larger value, getting away from the BF bound in the stable direction, the dual operator having now dimension $\Delta=2+Q_f>2$. 
%On the other hand, the perturbation of the mode in the probe approximation, even if feeling an $AdS_2$ spacetime in the IR, is effectively massless, and backreaction corrections to this cannot push perturbatively the mass beyond the BF bound for $AdS_2$.
%
%\vspace{0.5cm}
Therefore, we can speculate that if instabilities of the system were to be found they  arise as new solutions, presenting a spatially modulated phase or maybe something more complicated.

{\small
\section*{Acknowledgments}

We thank Ant\'on Faedo, Pau Figueras, Javier Mas, David Mateos, Andy O'Bannon and Domenico Seminara for useful discussions.

J.T. is supported by MEC FPA2010-20807-C02-02, by ERC StG HoloLHC - 306605 and by the Juan de la Cierva program of the Spanish Ministry of Economy.

\emph{F.B. and A.L.C. would like to thank the Italian students, parents, teachers and scientists for their activity
in support of public education and research.}
}

%%%%%%%%%%%%%%%%%%%%%%%%%%%%%%%%%%%%%%%%%
\appendix

\section{Expressions involved in the study of the perturbations}\label{app.perturbations}

By construction, the equations of the modes we focus on are those that vanish identically in our background. Specifically, these are the equations of motion for the field strengths $F^{(1)}_1$, $F^{(3)}_1$, $F^{(3)}_3$, $F^{(5)}_2$, $H^{(3)}_1$, $H^{(3)}_2$ and $d\AR$, which are a total of twelve equations so that our system may seem overdetermined (see \cite{Cotrone:2012um}). 
However, it is not difficult to show that the equation of motion for $F^{(3)}_3$ still vanishes identically, reducing the number of equations to nine and suggesting that the system is actually under-determined. This conclusion is also too naive. 
The reason is explained in \cite{Cotrone:2012um}, where it is shown that acting with the external derivative on the equations of motion for $F^{(5)}_2$ and $H^{(3)}_2$ gives rise to the equations of motion for $F^{(5)}_1$ and $d\cA_0$, respectively. The ultimate reason for this is that the modes $C^{(4)}_0$ and $B^{(2)}_0$ are nothing but St\"uckelberg scalars coupled to the vectors $C^{(4)}_1$ and $B^{(2)}_1$.\footnote{Actually, the St\"uckelberg scalar associated to the vector $B^{(2)}_1$ is a combination of the scalars $B^{(2)}_0$ and $\cA_0$.} The equation of motion for $F^{(5)}_1$ does not provide any new physical information, since it is not independent of the other equations of motion. On the other hand, the equation for $d\cA_0$ does provide a new, independent equation.

%%%%%

 \subsection{Dimension 4 scalar operator}

The first equation is the one for the fluctuation of the axion and at leading order ${\cal O}(\epsilon_*^0)$ is given by the equation of motion of a massless scalar in Schwarzschild-$AdS$
\be\label{eq.axionfluc}
{ C^{(0)}_0}'' + \partial_r \log \left[ r(r^4-r_h^4) \right] { C^{(0)}_0}' + \frac{r^4\omega^2}{(r^4-r_h^4)^2} { C^{(0)}_0} = 0 \ .
\ee

\subsection{Dimension 7 and 3 vector operators}
The second set consists of five different equations for the fluctuations of $A_{1}$, $C^{(4)}_1$ and $C_0^{(4)}$. 
At leading order in $\epsilon_*$ these are fluctuations of ${\cal N}=4$ SYM that can be considered independently of the backreaction of fundamental matter. 
It is convenient to express the vector fields in a different basis where we have a massive and a massless vector field dual to operators of dimension $\Delta_{\Sigma}=7$ and $\Delta_{\Delta}=3$ respectively,  which is achieved by taking
\be
 A_{1}=\frac{-\sqrt{2}\Delta+2\Sigma}{12} \ , \qquad  C^{(4)}_1= \frac{2\Delta+\sqrt{2}\Sigma}{12} \ .
\ee
Then the five equations for the fluctuations read
\begin{align}\label{eq.eq2}
\omega ( \omega \Delta_r- i \Delta_t ') & = 0  \ ,  \\ \label{eq.eq3}
3\omega \Delta_r +r \omega \Delta_r' - i (3\Delta_t' + r \Delta_t'')& = 0  \ ,  \\\label{eq.eq4}
-3i(r^4-r_h^4)\omega \Sigma_r +24 r^3 \Sigma_t +24 i r^3\omega  C^{(4)}_0 & \nonumber \\
- i (r^4-r_h^4) (r \omega \Sigma_r' - i (3 \Sigma_t'+r \Sigma_t''))  & = 0  \ ,  \\\label{eq.eq5}
(5r^8-6r^4r_h^4+r_h^8) \Sigma_r + i r^5 \omega \Sigma_t + r(r^4-r_h^4)^2 \Sigma_r' -r^5\omega^2  C^{(4)}_0 & \nonumber \\
- (5r^8-6r^4r_h^4+r_h^8)  {C^{(4)}_0}' - r(r^4-r_h^4)^2  {C^{(4)}_0}'' & = 0  \ ,  \\
-\sqrt{2} r^2 \omega^2 \Delta_r + (-48r^4+48r_h^2+2r^2\omega^2) \Sigma_r + ir^2 \omega  (\sqrt{2} \Delta_t'-2\Sigma_t') &\nonumber  \\ \label{eq.eq6}
+48(r^4-r_h^4)  {C^{(4)}_0}' & = 0  \ .
\end{align}
Notice also that $\Delta_t$ enters only via derivatives in the equations of motion, so we can always make a shift  to set the constant part of $\Delta_t$ to zero.

To solve \eqref{eq.eq2} and \eqref{eq.eq3} we just need to impose
\be\label{eq.Dradial}
\Delta_r = \frac{i}{\omega} \Delta_t' \ ,
\ee
where we are assuming $\omega\neq0$ in the following. With this plugged back into the equations, all dependence on the $ \Delta_{t,r}$ functions disappears, and one can algebraically solve for $\Sigma_r$ as well
\be\label{eq.Cradial}
\Sigma_r = \frac{24(r^4-r_h^4){ C^{(4)}_0}'-i r^2 \omega \Sigma_t' }{24(r^4-r_h^4)-r^2\omega^2} \ ,
\ee
and plugging \eqref{eq.Cradial} into \eqref{eq.eq4} and \eqref{eq.eq5} one gets the same differential equation involving zeroth, first and second derivatives of $\Sigma_t$ and $ C^{(4)}_0$. The solution to this equation is given by
\be\label{etaCdef}
 C^{(4)}_0 = \frac{i}{\omega} \Sigma_t + \eta_C \ ,
\ee
where $\eta_C$  satisfies
\be\label{eq.etaC}
\eta_C'' + \partial_r \log \left[\frac{r^3(r^4-r_h^4)}{24(r^4-r_h^4)-r^2\omega^2} \right] \eta_C' - \frac{24r^2(r^4-r_h^4)-r^4\omega^2}{(r^4-r_h^4)^2} \eta_C = 0 \ .
\ee
Before solving this equation of motion let us outline that the relations \eqref{eq.Dradial} and \eqref{eq.Cradial} are ambiguous. The fact that we cannot fix completely the five modes present in these equations of motion is reflecting the existence of a gauge freedom. In particular, two modes can be gauged away (this can be seen from the set \eqref{eq.eq2}--\eqref{eq.eq6} by noticing that there are two differential relations between these equations). 

\vspace{0.5cm}
To solve numerically the equation of motion for the fluctuation it is better to work with a  compact radial variable defined as $z=1/r$. In these coordinate,\footnote{In the rest of the appendix we will use $z$ radial coordinate in expressions necessary for the numeric integration and $r$ radial coordinate otherwise.} equation \eqref{eq.etaC} is solved near the boundary $z=0$ by
\begin{align}\label{eq.etaCUV}
\eta_C \approx & \frac{s_C}{z^4} + \frac{s_C\, \omega^2}{24 z^2} + \frac{s_C\, \omega^4}{1152} - r_h^4 s_C - \frac{s_C\, \omega^2}{24}r_h^4 z^2 - \frac{s_C\, \omega^4  (1728 r_h^4+\omega^4)}{442368}z^4 + v_C z^6 \\
& + \frac{\omega^2(-3317760 v_C+s_C \omega^2(262656r_h^8 + 1848 r_h^4 \omega^4+\omega^8))}{53084160} z^8 + \cdots \nonumber\\
& + \log[z] \left( \frac{s_C \, \omega^2 (147456r_h^8+1728 r_h^4\omega^4+\omega^8)}{2211840} z^6 \right. \nonumber\\ 
& \left. - \frac{s_C \, \omega^4 (147456r_h^8+1728 r_h^4\omega^4+\omega^8)}{35389440} z^8 + \cdots  \right) \nonumber \ ,
\end{align}
and we see that the asymptotic solutions in the UV behave as $z^{-4}$ and $z^{6}$, corresponding to the dual to a vector operator of dimension $\Delta=7$. 

In \cite{vanRees:2011fr} it was shown that for irrelevant operators the  coefficient multiplying the normalizable mode, proportional to $z^6$, is still holographically dual to the expectation value of the operator, whereas the non-normalizable $z^{-4}$ mode is dual to the source of the operator in the field theory.

Near the IR there are two independent solutions, one corresponding to a ingoing wave at the horizon\footnote{We are assuming here that the temperature is non-zero, $r_h\neq0$.} and one corresponding to an outgoing wave. Focusing in the ingoing wave we obtain
\be\label{eq.etaCIR}
\eta_C \approx (1-r_h z)^{-\frac{i \omega}{4r_h}} \eta_{C,h} \left[ 1+ \frac{3(128 r_h^3-32 i r_h^2 \omega+2 r_h \omega^2+i\omega^3)}{8\omega (2ir_h+\omega)} \left( z-\frac{1}{r_h}\right) + \cdots \right] \ ,
\ee
where $\eta_{C,h}$ is an undetermined normalization.

\vspace{0.5cm}
We consider now regularity at the horizon, which is obvious from the previous expression for $\eta_C$, but not evident for the physical modes $ C_0^{(4)}$ and $ \Sigma_t$. A quick look at the previous equations  shows that, assuming regularity for the time components, the only problems that can arise come from equation \eqref{eq.Cradial}. Using an ingoing wave regular solution for the time component of the gauge field, and the $\eta_C$ expansion \eqref{eq.etaCIR}, shows that to have a regular radial component of the vector field, $ \Sigma_{1r}$,  the fluctuation of the time components must vanish at the horizon, i.e.,
\be
 \Sigma_{t}  =  \# (r-r_h)^{1-\frac{i\omega}{4r_h}} + \cdots \ .
\ee
Imagine for a moment we pick  the St\"uckelberg gauge in which $ C^{(4)}_0=0$. Then from \eqref{etaCdef} we obtain that the time component  $ \Sigma_t$ has a finite value at the horizon (proportional to $\eta_{C,h}$) and the radial component would be irregular at the horizon, behaving like $ \Sigma_{r}\sim (r-r_h)^{-1}$. This is therefore an irregular gauge. From now on we will pick the regular, radial gauge\footnote{The same discussion will hold for the fluctuation of the NSNS potential below.}   $ \Sigma_{r}=0$. 
We obtain\footnote{Notice that the solution for $\Delta$, i.e., the field dual to the $\Delta_\Delta=3$ operator,  will not be needed; actually, this solution cannot be found uniquely since this is a pure gauge mode in the case of zero momentum. }
\be\label{eq.integral1}
 \Sigma_t  = \int dr \frac{24i(r^4-r_h^4)\omega }{24(r^4-r_h^4)-r^2\omega^2} \eta_C' \ , 
\ee
and performing an integral we obtain the complete function. When one does this integral a constant appears that must be fixed somehow. This constant is actually unphysical. In the equations of motion the terms proportional to time components of the massive vector fields without a derivative appear always in a certain combination with the corresponding St\"uckelberg scalar without derivative, and the constant shift of the time components can be compensated by the opposite shift in the  St\"uckelberg scalar. We can use this to set the constant of integration in the previous equations to zero.

The only physical parameter appearing in the equation of motion for the fluctuation $\eta_C$ is the temperature, which is encoded in the radius of the horizon $r_h = \pi T + {\cal O}(\epsilon_*)$. This means that at leading order in $\epsilon_*$ one can build the dimensionless ratio $\omega/r_h$ and describe all the finite temperature solutions at the same time, a change of temperature scale being compensated by a change in frequency.

\vspace{0.5cm}
From equation \eqref{eq.greens} we are interested in determining $V_{C\leftarrow \varphi_C}$ and $\varphi_C$. From the holographic correspondence the last identification corresponds just to the boundary condition $\varphi_C=s_C$, whereas the former must read $V_{C\leftarrow \varphi_C} \propto v_C$ \cite{vanRees:2011fr} when $s_C$ is non-zero. We are not interested in determining the exact proportionality factor, but in the position of the poles. Therefore, to produce the plots in figure \ref{fig.etaC} we solve equation \eqref{eq.etaC} performing a double shooting  from the IR and the UV with $s_C=1$ in the boundary conditions \eqref{eq.etaCUV} and \eqref{eq.etaCIR}, and varying $\eta_{C,h}$ and $v_C$ such that the solution and its derivative match smoothly in the bulk.\footnote{The numeric condition we have set for this smooth juncture is that the maximum difference between the function or its derivative at the matching point is less than $10^{-10}$.} Then
\be\label{eq.greenscomponents}
\frac{V_{C\leftarrow \varphi_C}}{\varphi_C}\propto \frac{v_C}{s_C} \ ,
\ee
and we will renormalize the Green's function such that the constant of proportionality is exactly one.
Actually, to study the two-point function we are interested in we should not focus just on $v_C/s_C$ but in the result coming from undoing the change of variables \eqref{eq.integral1}. Since we are evaluating an antiderivative (no extra constant must be considered)  it can be performed with the asymptotic series in the UV, to obtain
\be\label{eq.realCop}
\langle{\cal O}_\Sigma{\cal O}_\Sigma\rangle_R \propto \frac{v_C}{s_C} - \frac{r_h^8 \omega^2}{36}- \frac{r_h^4 \omega^6}{3072}- \frac{\omega^{10}}{5308416} \ .
\ee
The frequency powers correspond  to  contact terms (derivatives of Dirac's delta distributions when we Fourier-transform back). In particular they are analytic in the complex frequency plane. Our main focus is to determine the non-analytic behavior of the two-point function, so we may ignore these contact terms.

\subsection{Dimension 5 vector operator}

The third set of equations of motion contains three coupled equations. As in the previous subsection, at leading order in $\epsilon_*$ these are fluctuations of ${\cal N}=4$ SYM that can be considered independently of the backreaction of fundamental matter
\begin{align}
\label{eq.eq7} 8r(r^4-r_h^4)  B^{(2)}_{1r}- 2\sqrt{2} r (r^4-r_h^4)\delta {B^{(2)}_{0}}' - r^3 \omega (\omega  B^{(2)}_{1r}-i {B^{(2)}_{1t}}') & = 0 \ , \\
2\sqrt{2}ir^3\omega  B^{(2)}_{0}-3i(r^4-r_h^4)\omega  B^{(2)}_{1r} + 8r^3  B^{(2)}_{1t} - i\omega r(r^4-r_h^4){ B^{(2)}_{1r}}'    & \nonumber \\
\label{eq.eq8} -3(r^4-r_h^4){ B^{(2)}_{1t}}' - r (r^4-r_h^4){ B^{(2)}_{1t}}''   & = 0 \ , \\
 -r^5 \omega^2  B^{(2)}_{0} +2\sqrt{2}i r^5 \omega  B^{(2)}_{1t} +(5r^8-6r^4r_h^4+r_h^8) (2\sqrt{2}  B^{(2)}_{1r}  -  {B^{(2)}_{0}}' )  & \nonumber \\
\label{eq.eq9}  +r(r^8 - 2 r^4 r_h^4  +r_h^8)(2\sqrt{2} { B^{(2)}_{1r}}' - { B^{(2)}_{0}}'')    & = 0 \ .
\end{align}
As in the previous subsection, there is an algebraic solution for the radial component
\be\label{eq.Bradial}
 B^{(2)}_{1r} = \frac{2\sqrt{2} (r^4-r_h^4) {B^{(2)}_0}'  - i r^2 \omega  {B^{(2)}_{1t}}'  }{8(r^4-r_h^4)-r^2 \omega^2} \ ,
\ee
which, once plugged back into the equations of motion, identifies \eqref{eq.eq8} and \eqref{eq.eq9} in an equation with solution
\be\label{eq.B20}
 B^{(2)}_{0} = \frac{2\sqrt{2}i}{\omega}  B^{(2)}_{1t} +\eta_B\ ,
\ee
with $\eta_B$ satisfying
\be\label{eq.etaB}
\eta_B'' + \partial_r \log \left[\frac{r^3(r^4-r_h^4)}{8(r^4-r_h^4)-r^2\omega^2} \right] \eta_B' - \frac{8 r^2(r^4-r_h^4)-r^4\omega^2}{(r^4-r_h^4)^2} \eta_B = 0   \ .
\ee
As before, there is an ambiguity due to the gauge freedom (which can be seen in the set \eqref{eq.eq7}--\eqref{eq.eq8} by noticing the existence of a differential relation between the equations). Equation \eqref{eq.etaB} has as asymptotic solution in the UV the expansion
\begin{align}\label{eq.etaBUV}
\eta_B \approx  & \frac{s_B}{z^2} + \frac{s_B\, \omega^2}{16}-r_h^4 s_B \,z^2 + v_B z^4 + \left( - \frac{v_B\, \omega^2}{8} + \frac{7s_B\, \omega^4 (192 r_h^4+\omega^4)}{49152} \right) z^6 + \cdots \\
& + \log[z] \left( -\frac{s_B\, \omega^2 (102r_h^4+ \omega^4)}{768} z^4 + \frac{s_B\, \omega^4 (192 r_h^4+\omega^4)}{6144} z^6+ \cdots \right) \ , \nonumber
\end{align}
where the term proportional to the non-normalizable mode, $s_B$, is dual to the source and the term proportional to the normalizable mode, $v_B$, is related to the  vev of the dual, dimension $\Delta=5$ operator. Near the IR the ingoing wave solution is given by
\be\label{eq.etaBIR}
\eta_B \approx  (1-r_h z)^{-\frac{i\omega}{4r_h}} \eta_{B,h} \left[ 1 + \frac{128 r_h^3 - 32i r_h^2 \omega + 6 r_h \omega^2 + 3 i \omega^3}{8\omega (2i r_h+\omega)} \left( z -\frac{1}{r_h} \right) + \cdots \right]  \ , 
\ee
with $\eta_{B,h}$ a normalization.

The same discussion about regularity at the horizon that was performed for the $ \Sigma$ vector field can be performed in the present case, and we pick the radial gauge $ B_r^{(2)}=0$ to ensure regularity. Then we have the relation
\be
 {B^{(2)}_{1t}}  =\int dr \frac{2\sqrt{2} i (r^4-r_h^4) \omega}{8(r^4-r_h^4)-r^2\omega^2} \eta_B' \ ,\label{eq.integral2}
\ee
where the integral must be considered, once again, as an antiderivative.

\vspace{0.5cm}
Once again, the only physical scale at leading order in flavor backreaction is the temperature, and we can express the solution in terms of $\omega/r_h$. The results presented in figure \ref{fig.etaB} correspond to the identification
\be\label{eq.realBop}
\frac{V_{B\leftarrow \varphi_B}}{\varphi_B} = \frac{v_B}{s_B} \ ,
\ee
which can be obtained with a double shooting from the IR and UV with $s_B=1$ and the parameters $\eta_{B,h}$ and $v_B$ adjusted to match smoothly the solution and its derivative. The real two-point function component we are interested in must be obtained undoing the transformation \eqref{eq.integral2}, which gives us a result
\be
\langle{\cal O}_B{\cal O}_B\rangle_R \propto \frac{v_B}{s_B} - \frac{r_h^4 \omega^2}{8}- \frac{\omega^{6}}{1024} \ ,
\ee
and again the frequency powers correspond just to some contact terms that we can ignore since we want to focus on non-analytic terms. 

\subsection{Dimension 2 scalar operator}\label{subsec.A}
The  modes studied in the previous subsections enter as sources in the first order equation of motion for the fluctuation of the worldvolume scalar, $\delta{\cal A}_0$. Plugging the previous algebraic solutions we find that the equation of motion for the worldvolume scalar is solved by
\be
 {\cal A}_0 = -\frac{i}{\omega}  B^{(2)}_{1t}+ \eta_{\cal A} \ ,
\ee
with $\eta_{\cal A}$ satisfying
\begin{align}\label{eq.etaA}
\eta_{\cal A}'' + \partial_r \log\left[  \frac{(r^4-r_h^4)\sqrt{r^6+r_d^6}}{r^2} \right] \eta_{\cal A}' & \nonumber \\
+ \frac{r^4 \left( 8r(r^4-r_h^4)\sqrt{r^6+r_d^6} - 4r^4(r^4-r_h^4)+(r^6+r_d^6)\omega^2 \right) }{(r^6+r_d^6)(r^4-r_h^4)^2}\eta_{\cal A} & \nonumber\\
+ 2\sqrt{2} \frac{2r^{10}+2r^6 r_h^4 - r^4 r_d^6 + 5r_d^6 r_h^4}{r(r^6+r_d^6)(8(r^4-r_h^4)-r^2\omega^2)} \eta_B' + \sqrt{2} \frac{r^2(r^6+2r_d^6+2r^3\sqrt{r^6+r_d^6})}{(r^6+r_d^6)(r^4-r_h^4)} \eta_B & \nonumber \\
- 2i \frac{r^5 r_d^3 \omega}{(r^6+r_d^6)(24(r^4-r_h^4)-r^2\omega^2)} \eta_C' & = 0 \ .
\end{align}
Since ${\cal A}_0$ is dual to a dimension $\Delta=2$ operator the two independent solutions in the UV behave as $z^{2}$ and $z^{2}\log[z]$. In these limiting cases where the Brietenlohner-Freedman bound is saturated,  it is the coefficient of the most divergent term, namely the one with $z^2 \log[z]$, the dual to the source of the operator, whereas the coefficient of $z^2$ will be proportional to the expectation value \cite{Karch:2005ms}. 
Since $\eta_B$ and $\eta_C$ are dual to irrelevant operators that deform the UV of the theory, the asymptotic behavior is dominated by these fields \cite{vanRees:2011fr}. Using \eqref{eq.etaCUV} and \eqref{eq.etaBUV}, we find 
\begin{align}\label{eq.etaAUV}
\eta_{\cal A} \approx & - \frac{s_B}{2\sqrt{2} z^2} + \frac{\omega(16i s_C r_d^3-3\sqrt{2} s_B \omega)}{192} + v_{\cal A}\, z^2 \\
& + \frac{55 i r_d^3 s_C \omega^5+288( 4\sqrt{2} v_B + (-4 s_{\cal A} +\sqrt{2} s_B r_h^4 -4 v_{\cal A}) \omega^2)}{4608}z^4 + \cdots \nonumber \\
&  - \log[z] \left( s_{\cal A} z^2 + \frac{\omega^2 (-768 s_{\cal A} + 48 i s_C r_d^3 \omega^3 + \sqrt{2} s_B(192 r_h^4+\omega^4))}{3072} z^4 + \cdots \right) \nonumber \\
& + \log[z]^2 \left( - \frac{i\,r_d^3 \omega^3 s_C}{32}z^2 +  \frac{i\, r_d^3 \omega^5 s_C}{128} z^4 +  \cdots \right) \ .
\end{align}
The presence of the logarithm introduces a scale $z_0$ (which we fixed to 1) that enters in the counterterms one has to add to regularize the on-shell action. This affects the definition of the two-point function introducing contact terms (analytic terms in the momentum), which we are ignoring because we are interested only in the position of the poles of the retarded two-point functions, so our results will be insensitive to this scale. For the IR, the solution with the ingoing wave boundary conditions is given asymptotically by 
\begin{align}
\eta_{\cal A} \approx (1-r_h z)^{-\frac{i \omega}{4r_h}}  \Bigg[ &\eta_{{\cal A},h} + \Bigg(  \frac{r_h^2\left( 8ir_h(r_d^6+r_h^6)+ 2 \omega r_h^3 \sqrt{r_d^6+r_h^6} + (2 r_d^6+r_h^6)\omega  \right)}{\sqrt{2} (r_d^6+r_h^6)(2r_h-i\omega)\omega} \eta_{B,h}   \nonumber   \\
& +  \frac{-16 r_h^8 + 32 r_h^5 \sqrt{r_d^6+r_h^6} +6ir_d^6 r_h \omega-6ir_h^7\omega + 3 (r_d^6 +  r_h^6 )\omega^2}{8(r_d^6+r_h^6)(2r_h-i\omega)} \eta_{{\cal A},h}  \nonumber \\
& + \frac{r_d^3 r_h^5}{(r_d^6+r_h^6)(2r_h-i\omega)} \eta_{C,h} \Bigg) \left( z-\frac{1}{r_h}\right) + \cdots \Bigg] \ . \label{eq.etaAIR}
\end{align}

%In the radial gauge
%\be\label{eq.integral3}
% {\cal A}_0 = \eta_{\cal A} - \frac{i}{\omega}  B_{1t}^{(2)}\ .
%\ee
To integrate the equation, we perform again a double shooting from the IR and UV. Since we are interested in the two-point function associated to the operator of dimension $\Delta_{\cal A}=2$ we must make sure that we are sourcing only this operator. This implies that we have to integrate with conditions $s_{\cal A}=1$ and $s_B=s_C=0$. From this integration we can extract
\be
\frac{V_{{\cal A}\leftarrow \varphi_{\cal A}}}{\varphi_{\cal A}} = \frac{v_{\cal A}}{s_{\cal A}} \ .
\ee
The result for $v_{\cal A}/s_{\cal A}$ is reported in figure \ref{fig.etaA}.

\vspace{0.5cm}
The operators dual to $\eta_B$ and $\eta_C$ will source as well the operator dual to $\eta_{\cal A}$. This is seen from equation \eqref{eq.etaAUV}, where the presence of non-trivial $\eta_B$ or $\eta_C$ would trigger a non-trivial value of $v_{\cal A}$, even if we keep the boundary condition $s_{\cal A}=0$. We indeed observe this when we integrate the equations of motion with boundary conditions (a) $s_{\cal A}=s_B=0$ and $s_C=1$, giving
\be
\frac{V_{{\cal A}\leftarrow \varphi_C}}{\varphi_C} = \frac{v_{\cal A}}{s_C} \ ,
\ee
 or (b) $s_{\cal A}=s_C=0$ and $s_B=1$, giving
 \be
\frac{V_{{\cal A}\leftarrow \varphi_B}}{\varphi_B} = \frac{v_{\cal A}}{s_B} \ .
\ee

%%%%%%%%%%%%%

\end{document}